\documentclass[a4paper,11pt]{article}
\pdfoutput=1
\usepackage{jheppub}
\usepackage{slashed}
\title{Dynamics of a Stabilized Radion and Duality}
\author[a]{Zackaria Chacko,}
\author[a]{Rashmish K. Mishra,}
\author[a,b]{and Daniel Stolarski}
\affiliation[a]{Maryland Center for Fundamental Physics, 
Department of Physics, \\ 
University of Maryland, College Park, MD 20742-4111}
\affiliation[b]{Department of Physics and Astronomy,\\
Johns Hopkins University, Baltimore, MD 21218}

\abstract
{

We construct the effective theory of the graviscalar radion in the 
Randall-Sundrum scenario, taking into account effects arising from the 
stabilization of the extra dimension through the Goldberger-Wise 
mechanism. We explore the conditions under which the radion can remain 
light, and determine the corrections to its couplings to Standard Model 
(SM) states when the effects of stabilization are taken into account. We 
show that in the theories of interest for electroweak symmetry breaking 
that have a holographic dual, the presence of a light radion in the 
spectrum is not a robust prediction of the framework, but is in fact 
associated with mild tuning. We find that corrections to the form of the 
radion couplings to Standard Model particles arising from effects 
associated with brane stabilization are suppressed by the square of the 
ratio of the radion mass to the Kaluza-Klein scale. These corrections 
are small in theories where the radion is light, and are generally 
subleading, except in the case of couplings to the SM gluons and photon, 
when they can sometimes dominate. The AdS/CFT correspondence relates the 
radion in Randall-Sundrum models to the dilaton in theories where a 
strongly coupled conformal symmetry is spontaneously broken. We show 
that the discrepancies in the literature between the results for the 
dilaton and the radion can be traced to the omission of self-interaction 
terms that would otherwise dominate the potential for the 
Goldberger-Wise scalar near the infrared brane. In the dual picture, 
this corresponds to neglecting the corrections to the scaling behavior 
of the operator that breaks conformal symmetry when it grows large. With 
the inclusion of these self-interaction terms, we find good agreement 
between the results on the two sides of the correspondence.

}
\begin{document}
\maketitle
\flushbottom

\section{Introduction}
\label{sec:intro}

For almost four decades the Standard Model (SM) has enjoyed remarkable 
success in describing the strong and electroweak interactions. The 
recent discovery of a new particle with mass close to 125 
GeV~\cite{:2012gu,:2012gk}, if confirmed as the SM Higgs, will cement 
the SM as the correct effective description of nature at energies below 
the weak scale. However, there are good reasons to expect that the 
success of the SM will not continue. Within the SM there is no 
understanding of the stability of the weak scale under radiative 
corrections~\cite{Susskind:1978ms,'tHooft:1980xb}: the hierarchy problem. A solution to this problem requires 
new physics close to the weak scale that may be within reach of the 
Large Hadron Collider (LHC). Furthermore, the SM does not explain the 
hierarchy in the fermion masses or admit a dark matter candidate.

Among the many candidate theories that address the hierarchy problem, 
models based on the warped extra dimensional framework proposed by 
Randall and Sundrum~\cite{Randall:1999ee} are among the most promising. 
Randall-Sundrum (RS) models that incorporate a custodial isospin 
symmetry~\cite{Agashe:2003zs,Agashe:2006at} can be consistent with 
precision electroweak measurements even if the Kaluza-Klein (KK) scale 
is as low as a few TeV, allowing the hierarchy problem to be addressed. 
This has allowed calculable, holographic realizations of 
technicolor~\cite{Csaki:2003zu}, and of the Higgs as a 
pseudo-Nambu-Goldstone boson (pNGB)~\cite{Contino:2003ve,Agashe:2004rs}, 
to be constructed.  The RS framework naturally accommodates new 
approaches to the flavor problem of the SM~\cite{Grossman:1999ra, 
Gherghetta:2000qt, Agashe:2004cp}, 
and to 
the dark matter puzzle~\cite{Agashe:2004ci,Agashe:2004bm,Medina:2011qc}. 
Apart from these attractive theoretical features, RS models have very 
distinctive collider 
signatures~\cite{Agashe:2006hk,Contino:2006nn,Lillie:2007yh}, which may 
help in distinguishing them from other theories.

In RS models, fluctuations in the inter-brane separation are associated 
with a graviscalar field, the radion~\cite{Randall:1999ee}. Several authors have studied the 
dynamics of the radion in this framework~\cite{Goldberger:1999uk} and determined its couplings to 
the SM states, both in the case when the SM fields are localized to the 
infrared (IR) brane~\cite{Csaki:1999mp,Goldberger:1999un, Giudice:2000av, Csaki:2000zn} and the 
case when they are in the bulk~\cite{Csaki:2007ns, Rizzo:2002pq}.  In 
the original RS construction, the radion couplings are fixed by $5D$ diffeomorphism, but it
is massless, corresponding to 
the fact that the distance between the two branes is a free parameter. 
Once the geometry is stabilized using, for example, the Goldberger-Wise (GW) 
mechanism~\cite{Goldberger:1999uk}, the inter-brane separation is fixed 
and the radion acquires a mass. In general, we expect corrections to the 
radion couplings arising from effects associated with the dynamics that 
stabilizes the geometry. These corrections, which may be significant, 
have not been taken into account in previous analyses.

The AdS/CFT correspondence~\cite{Maldacena:1997re, Witten:1998qj,Gubser:1998bc,Klebanov:1999tb}, relates the original RS two brane 
model to strongly coupled large $N_c$ gauge theories that exhibit a 
conformal symmetry which is broken spontaneously~\cite{ArkaniHamed:2000ds,Rattazzi:2000hs}. The brane localized SM 
Higgs of the RS model is identified with a composite of the strong 
dynamics on the four dimensional side of the correspondence~\cite{ArkaniHamed:2000ds,Contino:2003ve,Agashe:2004rs}. The low 
energy spectrum of any theory where conformal symmetry is realized 
non-linearly contains a massless scalar, the 
dilaton~\cite{Salam:1970qk,Isham:1970gz,Isham:1971dv,Zumino:1970ab,Ellis:1970yd,Ellis:1971sa}, 
which can be thought of as the Nambu-Goldstone boson (NGB) associated 
with the breaking of scale invariance. In the correspondence, the 
massless radion of the RS model is identified with the 
dilaton~\cite{Rattazzi:2000hs,ArkaniHamed:2000ds}. The mechanism that stabilizes the brane 
separation, and thereby generates a mass for the radion, corresponds to 
the presence of explicit conformal symmetry violating effects in the 
strongly coupled theory.

Several authors have studied the dynamics of the dilaton in the context 
of theories where strong conformal dynamics constitutes the ultraviolet (UV) 
completion of the physics that drives electroweak symmetry breaking~\cite{Rattazzi:2000hs,Goldberger:2007zk,Fan:2008jk,Vecchi:2010gj,RattazziPlanck:2010,Chacko:2012sy,Bellazzini:2012vz,Abe:2012eu,Coradeschi:2013gda}. The 
mass of the dilaton~\cite{RattazziPlanck:2010, Chacko:2012sy,Bellazzini:2012vz,Coradeschi:2013gda} and its 
couplings to SM states, incorporating conformal 
symmetry violating effects~\cite{Chacko:2012sy,Bellazzini:2012vz}, have been determined. However, at present 
there is only very imperfect agreement between the results in the 
literature for the dilaton mass on the CFT side of the correspondence 
and the radion mass in RS models. The reasons for this discrepancy, and 
its resolution, remain to be understood.

In this paper we construct the effective theory of the radion in RS 
models, taking particular care to include effects arising from the 
dynamics that stabilizes the geometry. We explore the conditions under 
which the radion can remain light and determine the form of its 
couplings to Standard Model (SM) states. We show that the mass of the 
radion can naturally lie below the Kaluza-Klein scale if all the 
parameters in the potential for the GW scalar lie below their natural 
strong coupling values. Alternatively, the spectrum of states will 
include a light radion if the tension of the IR brane is tuned to lie 
close to its value in the original RS model. The original RS setup where 
the radion is exactly massless may be thought of as an extreme case of 
this fine-tuning. Since neither of these conditions is in general 
expected to be satisfied in the theories of interest for electroweak 
symmetry breaking that have a holographic dual, the presence of a light 
radion in the spectrum is not a generic prediction of this framework, 
but is instead associated with mild tuning. We find that corrections to 
the form of the radion couplings to SM fields arising from effects 
associated with brane stabilization are suppressed by the square of the 
ratio of the radion mass to the Kaluza-Klein scale, and are small in 
theories where the radion is light. These corrections are generally 
subleading, except in the case of radion couplings to the SM gluons and 
photon localized on the IR brane, when they can sometimes dominate.

Holography decrees that the physics of the radion correspond to that of 
the dilaton in the dual description. We find good agreement between our 
results for the radion, and those in the literature for the dilaton. We 
are also able to arrive at a holographic interpretation of our results 
for the radion. As part of our analysis, we resolve the discrepancy in 
the literature between the results for the dilaton mass on the CFT side 
of the correspondence, and the radion mass in RS models. In theories 
where the dilaton mass lies below the scale of strong dynamics, the mass 
squared of the dilaton is known to depend on the scaling dimension of 
the operator $\mathcal{O}$ that leads to the breaking of conformal 
symmetry. Near the breaking scale where this operator is large, its 
scaling behavior differs significantly from that at the fixed 
point. We trace the reason for the discrepancy between the expressions 
for the radion mass on the two sides of the correspondence to the fact 
that this effect has not previously been systematically taken into 
account in RS models. We show that incorporating this effect corresponds 
to the addition of non-linear terms in the bulk potential for the GW 
scalar. Once this effect is taken into account, we obtain good agreement 
between the AdS and CFT sides of the correspondence.

The organization of rest of the paper is as follows. In 
section~\ref{sec:rad-mass}, we consider the effect of stabilization of 
the extra dimension by a GW scalar $\Phi$ and construct the low energy 
effective theory. In section~\ref{sec:rad-coupling} we determine the 
couplings of the radion to SM fields on the visible brane taking the 
effects of stabilization into account. In section~\ref{sec:dual-pic}, we 
relate the obtained results to the dual picture making use of the 
AdS/CFT dictionary. We conclude in section~\ref{sec:conclusion}. Various 
technical details are given in the appendices.

\section{Stabilization of the Extra Dimension}
\label{sec:rad-mass}

In this section we construct the low energy effective theory of the 
radion incorporating the effects of the GW stabilization mechanism. Our 
primary objective will be to determine the dependence of the radion mass 
on the parameters of the extra dimensional theory. Following Goldberger 
and Wise we consider a scenario where a $5D$ scalar field $\Phi$ 
acquires a vacuum expectation value (VEV) whose value depends on the 
location in the extra dimension~\cite{Goldberger:1999uk}. However, we deviate from the original 
GW construction in that we allow the IR brane tension to be 
detuned away from its fine-tuned value in the original RS model, and also 
include non-linear self interaction terms for the GW field $\Phi$ in the 
bulk. As we shall see, this alters the parametric dependence of the 
radion mass. We will later elaborate on the significance of this by 
invoking the dual picture in section~\ref{sec:dual-pic}.

\subsection{Mass of the Radion}
\label{sec:radion_mass}

We begin by considering the $5D$ gravity action in the absence of any 
stabilization mechanism. 
 \begin{align}
\mathcal{S}_{GR}^{5D}
&=
\int d^4x\int_{-\pi}^{\pi}d\theta\;
\left[
\sqrt{G}\Big(-2M_5^3\mathcal{R}[G]-\Lambda_b\Big)
-\sqrt{-G_h}\delta(\theta)T_h-\sqrt{-G_v}\delta(\theta-\pi)T_v
\right], 
\label{eq:1.1}
 \end{align}
 where $M_5$ is the $5D$ Planck mass, $\Lambda_b$ is the bulk 
cosmological constant and $T_h$, $T_v$ are the brane tensions at the hidden 
(UV) and visible (IR) brane respectively. We are using the $(+,-,-,-,-)$ 
sign convention for the $5D$ metric. The classical solution for the 
geometry is a slice of AdS$_5$ compactified on a circle $\mathcal{S}_1$ with $\mathcal{Z}_2$ orbifolding. The background metric $G_{MN}$ is then given as 

 \begin{eqnarray}
ds^2&=e^{-2kr_c|\theta|}\eta_{\mu\nu}dx^\mu dx^\nu-r_c^2\:d\theta^2 
\qquad-\pi\leq\theta\leq\pi\; .
\label{eq:metric}
 \end{eqnarray}
 To obtain a static solution of this form we have set $\Lambda_b/k= 
T_v=-T_h=-24M_5^3k$, while $r_c$ is an arbitrary constant. To parametrize 
the light degrees of freedom, we promote $r_c$ and $\eta_{\mu\nu}$ to 
dynamical fields by letting $r_c\rightarrow r(x)$ and 
$\eta_{\mu\nu}\rightarrow g_{\mu\nu}(x)$, which are associated with the 
radion and the graviton fields respectively. After this replacement in 
the metric, we plug the metric back in the action and integrate over the 
extra dimension to obtain the effective theory for the $4D$ graviton 
$g_{\mu\nu}(x)$ and the canonically normalized radion field 
$\varphi(x)$, see 
appendix~\ref{app:a} for details. 
Up to exponentially small corrections, this is given by
 \begin{align}
\mathcal{S}_{GR}^{4D}
&=
\frac{2M_5^3}{k}\int d^4x\sqrt{-g}\:\mathcal{R}[g_{\mu\nu}]
+
\int d^4x\sqrt{-g}\;
\Big(\frac12\partial_\mu\varphi\partial^\mu\varphi
-V(\varphi)\Big) \; .
\label{eq:1.3}
 \end{align}
 From this we see that the $4D$ Planck mass is given by $M_4^2=M_5^3/k$. The canonically normalized field $\varphi(x)$ is related to $r(x)$ as 
 \begin{eqnarray}
\varphi(x)= F e^{-k \pi r(x)} \; ,
\label{eq:phi-def}
 \end{eqnarray} 
 where $F$ is defined as $F = \sqrt{24M_5^3/k}=\sqrt{24}\,M_4$. The potential generated 
for the radion field from the $5D$ gravity sector is given by
 \begin{align}
V_{GR}(\varphi)
&=\frac{\varphi^4}{F^4}\Big(T_v-\frac{\Lambda_b}{k}\Big) \; ,
\label{eq:1.4}
 \end{align}
 and vanishes identically for the tuned values of $\Lambda_b$ and $T_v$ 
in the original RS solution.

This naive procedure of promoting $\eta_{\mu\nu}$ and $r_c$ to dynamical 
fields leads to a parametrization of the light modes that does not solve 
the linearized Einstein equations~\cite{Charmousis:1999rg, Csaki:2000zn}. One may 
therefore worry that the low energy effective field theory (EFT) could 
get large corrections from the heavy fields.  However, this turns out 
not to be a concern, as shown in~\cite{Luty:2000ec}. A general 
parametrization of the fluctuations that includes both heavy and light 
modes can be written as
 \begin{align} 
ds^2=
e^{-2kr(x)|\theta|}\Big[g_{\mu\nu}(x)+ H_{\mu\nu}(x,\theta)\Big]dx^\mu dx^\nu 
+2H_{\theta\mu}(x,\theta)dx^{\mu}d\theta
-r^2(x)\Big[1+H_{\theta\theta}(x,\theta)\Big]\:d\theta^2 \; .
\label{eq:} 
 \end{align}
 Here $g_{\mu\nu}(x)$ and $r(x)$ are again the light fields, while 
$H_{\mu\nu}(x, \theta)$, $H_{\theta\mu}(x, \theta)$ and 
$H_{\theta\theta} (x, \theta)$ parameterize the heavy Kaluza-Klein (KK) 
excitations of the $5D$ graviton after reducing to the four dimensional 
theory. This naive parametrization of the light degrees of freedom does 
not lead to large corrections from the heavy fields in the low energy 
effective theory as long there are no tadpole terms in the heavy fields 
with one derivative~\cite{Luty:2000ec}. By $4D$ Lorentz invariance, such a 
tadpole must multiply $H_{\mu\theta}$, but an explicit computation shows 
that $H_{\mu\theta}$ vanishes identically for the metric that solves the 
linearized Einstein equation~\cite{Charmousis:1999rg,Csaki:2000zn}. This 
ensures that our naive parametrization of light fields will indeed 
generate the correct low energy effective theory up to corrections that 
are suppressed by powers of the KK scale. This conclusion is not altered 
by the inclusion of the GW field.

In order to solve the hierarchy problem, the warp factor in the 
background geometry must be exponentially large. Defining $f = 
\left<\varphi\right>=Fe^{-kr_c\pi}$, we require a stable solution such 
that $f \ll F$ . The radion potential in Eq.~\eqref{eq:1.4} generated from 
the gravitational part of the action does not allow for such a 
possibility dynamically, unless fine-tuned to vanish. The VEV of 
$\varphi$ either vanishes or is driven to infinity depending on the sign 
of the coefficient of the quartic. In the original RS solution to the 
hierarchy problem, this issue is addressed by setting the quartic 
coefficient to zero by tuning the tension of the visible brane such that 
$T_v= - 24M_5^3k = \Lambda_b/k$. The cosmological constant must also be 
tuned to zero by adjusting the tension of the hidden brane, $T_h = 
24M_5^3k = - \Lambda_b/k$.

In a realistic RS scenario, a mechanism is needed to stabilize the 
geometry and give the radion a mass. Such a mechanism was proposed by 
Goldberger and Wise~\cite{Goldberger:1999uk}. In this construction, a 
massive $5D$ field $\Phi$ is sourced at the boundaries and acquires a 
VEV whose value depends on the location in the extra dimension. After 
integrating over the extra dimension, this generates a potential for the 
radion field in the low energy effective theory. In the original GW 
construction, the quartic potential for the radion in Eq.~\eqref{eq:1.4} was 
tuned to zero, and only the dynamics of the scalar field $\Phi$ 
contributed to the radion potential.  Here we allow the quartic 
generated by the gravitational potential to be non-zero, and also 
include self-interaction terms in the potential for the GW scalar in the 
bulk. The effect of these additional terms, which are important in 
theories with a holographic dual, is to alter the parametric dependence 
of the radion mass.

We now construct the low energy effective theory for the radion that 
emerges after stabilization using the GW mechanism. The action for the 
GW scalar $\Phi$ is given by
 \begin{align}
\mathcal{S}_{GW}
&=\int d^4x\:d\theta\:
\left[
\sqrt{G}\:\left(
\frac12G^{AB}\partial_A\Phi\partial_B\Phi
-V_b(\Phi)
\right)
-\sum_{i=v,h}\delta(\theta-\theta_i)\:\sqrt{-G_i}\:V_i(\Phi)
\right] \; .
\label{eq:2.1}
 \end{align}
 Here $G_h$ and $G_v$ are the induced metrics at the brane locations, 
$\theta_h=0$ and $\theta_v=\pi$. The corresponding brane potentials are 
given by $V_h$ and $V_v$ respectively, while $V_b$ is the potential in 
the bulk. In our parametrization, the radion does not couple to the 
hidden brane at $\theta_h = 0$. We therefore do not specify the form of 
$V_h$, but simply require that it fixes $\Phi$ at $\theta=0$ to be 
$k^{3/2}v$ and does not contribute to the hidden brane tension. Such a 
requirement can be easily arranged, and is in fact the one used in the 
original GW proposal.

On the visible brane, we consider a simple potential for $\Phi$ of the form
 \begin{eqnarray}
V_v(\Phi)&=&2k^{5/2}\alpha\:\Phi \; ,
\label{eq:vis_brane}
 \end{eqnarray}
 where $\alpha$ is a dimensionless number. The choice of a linear 
potential for $\Phi$ on the IR brane is not special and is expected to present if $\Phi$ is not charged under any symmetries. Considering a 
more general brane potential does not alter our final conclusions.

The potential for the GW scalar $\Phi$ in the bulk has the general form
 \begin{eqnarray}
V_b(\Phi) = \frac12m^2\Phi^2+\frac{1}{3!}\eta\Phi^3+\frac{1}{4!}\zeta\Phi^4+...\;\;\;.
\label{eq:bulk_pot}
 \end{eqnarray}	
 Since we are in AdS space, the mass squared parameter $m^2$ for the GW 
scalar $\Phi$ can be negative without giving rise to instabilities as 
long as the condition $m^2/k^2+4>0$ is 
satisfied~\cite{Breitenlohner:1982jf}. The theories of interest for 
electroweak symmetry breaking are characterized by a large hierarchy 
between the scales associated with the UV brane and the IR brane, ${\rm 
exp} (- k \pi r_c) \ll 1$. The size of the hierarchy is determined by 
the potential for the GW scalar, which fixes the brane spacing $r_c$. A 
large hierarchy can be obtained if the mass term in the bulk potential 
for the GW field is small in units of the inverse curvature $k$. As 
first observed by Goldberger and Wise~\cite{Goldberger:1999uk}, a value of $|m^2/k^2|$ of order a 
tenth suffices to generate the hierarchy between the Planck and weak 
scales.

In the class of RS models that possess a holographic dual, the 
parameters in the $5D$ gravity theory and in the potential for the GW 
scalar are related to parameters in the dual CFT. The AdS/CFT 
correspondence associates the GW field $\Phi$ with a scalar operator 
$\mathcal{O}$ that deforms the dual CFT. Our primary focus is on the 
duals of $4D$ theories where a nearly marginal deformation 
$\mathcal{O}$, although small in the UV, grows large in the IR, 
resulting in the breaking of the conformal symmetry. This dictates our 
choice of the values of the parameters in the $5D$ theory. The mass of 
$\Phi$ is related to the scaling dimension of $\mathcal{O}$ at the fixed 
point, and $|m^2/k^2| \ll 1$ corresponds to the operator $\mathcal{O}$ 
being close to marginal. We choose $m^2 < 0$ to obtain a solution for 
$\Phi$ which grows in the IR. The dimensionless parameter $v$, which 
determines the value of $\Phi$ at the UV brane is also chosen to be 
small, allowing us to work to linear order in $v$. This corresponds to 
the operator $\mathcal{O}$ being a small deformation in the UV. The 
other parameters are taken to be of order their natural strong coupling 
values (see appendix~\ref{app:NDA} for a description of how to estimate 
the natural values of parameters), but with an important caveat. In our 
analysis, we neglect the back-reaction of the scalar field dynamics on 
the metric. In order for this approximation to be self-consistent, the 
contribution of the GW field to the total energy density at any point 
must be smaller, by a factor of at least a few, than that of the 
cosmological constant. For this to be the case the detuning of the IR 
brane tension away from that in the original RS model must be smaller, 
by a factor of order a few, than its natural strong coupling value. The 
value of the parameter $\alpha$, which controls the value of $\Phi$ near 
the IR brane, must also be taken to be smaller than its natural value by 
a similar factor. We do not expect our general conclusions to be 
affected by the fact that we are working in a limit where the back-reaction is small.

Given the action, we can solve for the classical field configuration 
$\Phi(\theta)$ in the RS background. The equation satisfied by $\Phi$ in 
the bulk with the boundary conditions resulting from brane potentials is 
given by
 \begin{eqnarray}
&&\partial_\theta^2\Phi
-4kr_c\partial_\theta \Phi
-r_c^2V_b'(\Phi)=0
\nonumber\\
&&\theta=0\qquad:\qquad \Phi=k^{3/2} v
\nonumber\\
&&\theta=\pi\qquad:\qquad \partial_\theta\Phi=-\alpha k^{3/2} kr_c\;\;.
\label{eq:PhiEOM}
 \end{eqnarray}
 The classical equation is a second order differential equation and is 
not simple to solve analytically for a general bulk potential. If, 
however, $4kr_c \gg 1$ and certain conditions are satisfied by the 
parameters in the bulk potential, then an approximate analytic solution 
may be obtained~\cite{RattazziPlanck:2010} using the methods of 
singular perturbation theory~\cite{benOrz}. The key observation is that 
in the limit $4kr_c \gg 1$, the bulk equation of motion in 
Eq.~\eqref{eq:PhiEOM} possesses solutions in different regions of 
$\theta$ that have the property that one of the terms in 
Eq.~\eqref{eq:PhiEOM} is small compared to the other two. In particular, 
there are self-consistent solutions for $\Phi$ in the bulk that satisfy 
the boundary condition on the UV brane such that the 
$\partial_\theta\Phi$ term and $V_b'$ term are parametrically larger 
than the $\partial_\theta^2\Phi$ term. Within this approximation, the 
equation effectively becomes first order and is readily solved. The full 
solution to the equation displays boundary layer formation very near the 
$\theta=\pi$ boundary where the $\partial_\theta^2\Phi$ term cannot be 
self-consistently dropped. However, in this regime it is self-consistent 
to drop the potential term in favor of the $\partial_\theta\Phi$ and 
$\partial_\theta^2\Phi$ terms.

It follows that two independent approximate solutions to the second 
order differential equation can be obtained in the following way: once 
by balancing the $\partial_\theta\Phi$ term against the potential term, 
and once by balancing the $\partial_\theta\Phi$ term against the 
$\partial_\theta^2\Phi$ term. We will refer to these two 
equations (solutions) as the outer region (OR) and boundary region (BR) 
equations (solutions) respectively. The OR solution holds in the bulk, 
while the BR solution holds close to $\theta=\pi$ where a boundary layer 
is formed, as the name suggests. The thickness of the boundary layer is 
of the order $\sim 1/4kr_c$. More specifically, the equations 
and their domains of validity\footnote{For the remainder of this work, we give solutions for $0 \leq \theta \leq \pi$. Because of the ${\cal Z}_2$ orbifold symmetry, this is sufficient to reconstruct the solution for the entire space.} are,
 \begin{eqnarray}
{\rm OR}:\qquad\;\;
\frac{d\Phi}{d\theta}&=& -\frac{r_c}{4k}V_b'(\Phi)
\qquad\;\;\left(0\leq\theta\lesssim\pi-\frac{1}{4kr_c} \right)
\nonumber\\
{\rm BR}:\qquad
\frac{d^2\Phi}{d\theta^2}&=&4kr_c\frac{d\Phi}{d\theta}
\qquad\qquad\left(\pi-\frac{1}{4kr_c}\lesssim\theta\leq\pi\right) \; .
\label{eq:}
 \end{eqnarray} Notice that the BR solution is independent of the choice 
of potential $V_b$ and can be readily solved. After applying the 
boundary condition at $\theta = \pi$, it is given by
 \begin{eqnarray} 
\Phi_{BR}(\theta)= -\frac{k^{3/2}\alpha}{4}e^{4kr_c(\theta-\pi)} +C\; .
\label{eq:} 
 \end{eqnarray} 
 The yet unspecified constant $C$ is determined by requiring the BR 
solution to be consistent with the OR solution, using asymptotic 
matching to the OR solution~\cite{benOrz}. This allows us to construct a smooth solution 
that is a good approximation to each of the two solutions in the 
appropriate region. It is clear that the BR solution 
is exponentially suppressed in the region $0\leq \theta \lesssim 
\pi-1/4kr_c$, but becomes important in the region 
$\pi-1/4kr_c\lesssim\theta \leq \pi$, thereby justifying the 
approximations made. Different choices of $V_b$ change the OR solution, 
and therefore change the constant $C$ in the BR solution. The complete 
solution for $\Phi$ exhibits the universal feature of 
boundary layer formation close to $\theta=\pi$. As we shall see, this
general characteristic is to be expected from the point of view of the 
holographic dual theory.

The potential for the GW scalar $\Phi$ in the bulk takes the form 
 \begin{eqnarray} 
V_b(\Phi) = \frac12m^2\Phi^2+\frac{1}{3!}\eta\Phi^3+\frac{1}{4!}\zeta\Phi^4+... \;\;\;.
\label{eq:bulk_pot} 
 \end{eqnarray} 
We can then write the OR equation in a holographically suggestive form 
\begin{eqnarray} 
\frac{d\,\log\Phi}{d\,(kr_c\theta)}= -\frac{m^2}{4k^2} 
-\frac{\eta}{8\sqrt{k}}\:\frac{\Phi}{k^{3/2}} 
-\frac{\zeta\:k}{24}\:\frac{\Phi^2}{k^3}+... \;\;\;.
\label{eq:full-OR-equation} 
 \end{eqnarray} 
 In what follows we restrict ourselves to specific forms of the bulk 
potential for the GW field. In particular, we will first study the case 
when $V_b(\Phi)$ consists of just a mass term, and then consider the case of 
a cubic self-interaction term. For each theory, we will describe the conditions on the parameters of the bulk potential for our approximation to be valid.
We will then obtain the 
OR solution and study the resulting radion potential. These 
special cases will suffice to determine the general features of the 
solution.

\subsubsection*{Massive GW Scalar with no Bulk Interactions}

\begin{figure}[ht]%
\centering
\includegraphics[width=\textwidth]{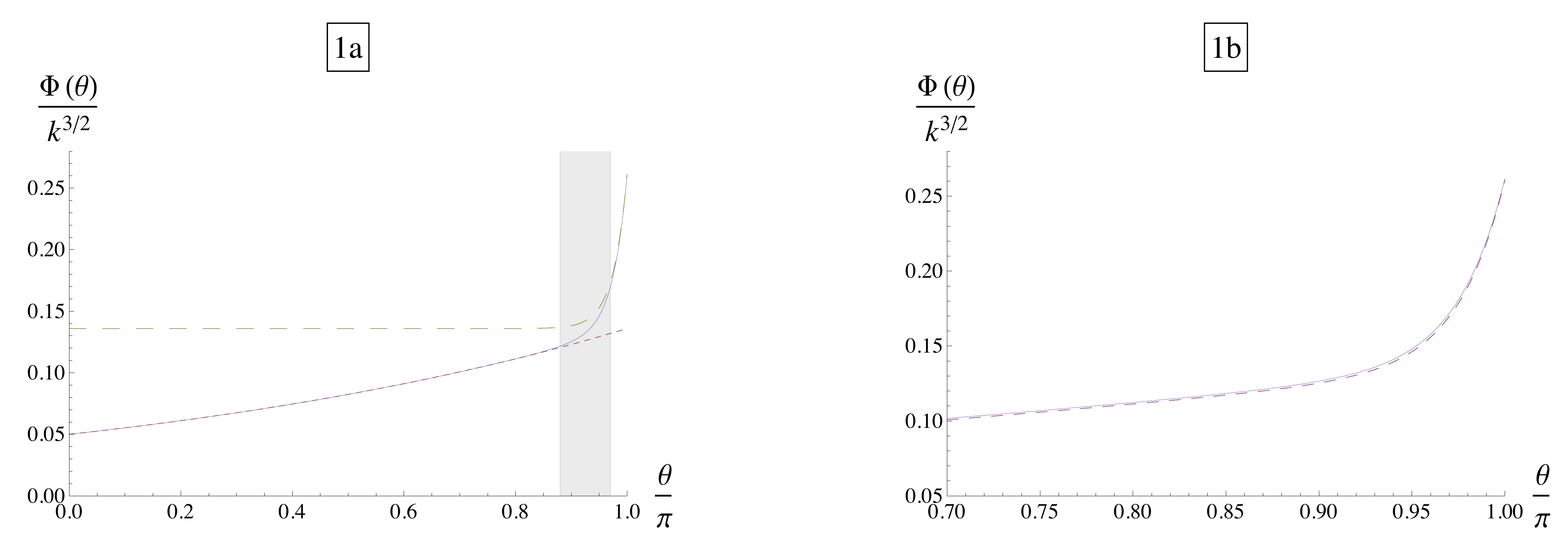}%
\caption{\;\textbf{1a}: The approximate solution (solid line) matches well with the BR solution (large dotted) and the OR solution (small dotted) for $\epsilon=-0.1,\,k\pi r_c=10$. We have also taken $v=0.05$ and $\alpha=-0.5$. The shaded region separates the boundary region on its right from the outer region on its left. Asymptotic matching is done in the shaded region. \textbf{1b}: The approximate solution (dotted) agrees well with the exact solution (solid) for the same parameter values, and we show the agreement near the $\theta=\pi$ boundary.}%
\label{fig:1}%
\end{figure}

We begin by considering the case when the potential for $\Phi$ is dominated 
by the mass term, and the higher powers of $\Phi$ can be neglected, 
 \begin{eqnarray}
V_b(\Phi)=\frac 12m^2\Phi^2\;.
 \label{eq:}
 \end{eqnarray} 
 The large hierarchy between the Planck and weak scales implies that 
$e^{-kr_c\pi}$ is a very small number. To obtain this hierarchy we 
require the bulk mass squared being small in units of 
the inverse curvature, namely $\epsilon = m^2/4k^2$ is a small number, 
$|\epsilon| \ll 1$.  We also take $e^{-\epsilon kr_c\pi}$ to be an 
$\mathcal{O}(1)$ number. These conditions are also sufficient for our boundary layer analysis to be valid.   The OR equation in this case takes the form
 \begin{eqnarray}
\frac{d \log \Phi}{d (kr_c\theta)}= -\frac{m^2}{4k^2}=-\epsilon \; .
\label{eq:}
 \end{eqnarray}
 Applying the boundary condition at $\theta=0$, the solution is
 \begin{eqnarray}
\Phi_{OR}(\theta)=k^{3/2}ve^{-\epsilon kr_c\theta}\; .
\label{eq:massive-phi-OR}
 \end{eqnarray}
 Given this OR solution, the constant $C$ in the BR solution is fixed 
uniquely by requiring asymptotic matching to the OR solution, resulting 
in
 \begin{eqnarray}
\Phi_{BR}(\theta)=-\frac{k^{3/2}\alpha}{4}e^{4kr_c(\theta-\pi)}+k^{3/2}ve^{-\epsilon kr_c\pi}\; .
\label{eq:}
 \end{eqnarray}
 Using the form of the OR and BR solutions, a smooth approximate 
solution for $\Phi$ that is valid in the entire region 
$0\leq\theta\leq\pi$ and that matches on to both $\Phi_{BR}$ and 
$\Phi_{OR}$ is given by
 \begin{eqnarray}
\Phi_{\rm approx}(\theta)=-\frac{k^{3/2}\alpha}
{4}e^{4kr_c(\theta-\pi)}+k^{3/2}ve^{-\epsilon kr_c\theta}\;.
\label{eq:massive-phi-complete}
 \end{eqnarray}
 For this choice of the bulk potential, an exact expression for $\Phi$ 
can be obtained (see appendix~\ref{app:b}), which matches very well with 
the approximate solution. Figure~\ref{fig:1} indicates the matching of 
the BR and IR solutions to the approximate solution (1a), and shows the 
agreement of the approximate solution with the exact solution (1b).

To calculate the resulting contribution to the radion potential, we 
promote $r_c$ to a dynamical field and insert the complete solution for 
$\Phi(\theta)$ with this replacement back into the action. Integrating 
over the extra dimension then generates the contribution to the radion 
potential (see appendix~\ref{app:b} for details). To leading order in 
$v$ and $\epsilon$, the resulting potential takes the form
 \begin{align}
V_{GW}(\varphi)&=
k^4\left(\frac{\varphi}{F}\right)^{4}
\left[
2\alpha v\left(\frac{\varphi}{F}\right)^{\epsilon}
-\frac{\alpha^2}{4}
\right]
 \; .
\label{eq:2.5}
 \end{align} 
 Including the contribution from the gravity part of the action in Eq.~\eqref{eq:1.4}, the full potential for the radion is given to leading 
order in $v$ and $\epsilon$ as
 \begin{align}
V(\varphi)&=
V_{GR}(\varphi)+V_{GW}(\varphi)\nonumber\\
&=
2k^4\alpha v\left(\frac{\varphi}{F}\right)^{4+\epsilon}
+\Big(T_v-\frac{\Lambda_b}{k}-\frac{k^4\alpha^2}{4}\Big) \left(\frac{\varphi}{F}\right)^4\; .
\label{eq:rad-potential-1}
 \end{align} 
 We can define $\tau=\left(T_v-\Lambda_b/k-k^4\alpha^2/4\right)/k^4$ 
which represents the detuning of the brane tension away from the 
original RS solution. The first two terms in $\tau$ are purely 
gravitational, while the last term comes from the dynamics of the GW 
field on the visible brane and receives corrections higher order in 
$\epsilon$. With this, the minimum of the potential in 
Eq.~\eqref{eq:rad-potential-1} occurs at $\left<\varphi\right>=f$ where 
the condition
 \begin{eqnarray}
\left[\frac{f}{F}\right]^\epsilon\equiv e^{-\epsilon kr_c\pi}
&=&-\frac{\tau}{2\alpha v}+\mathcal{O}(\epsilon)
\label{eq:2.7}
\end{eqnarray}
 is satisfied. For small $\epsilon$, an exponential hierarchy can be 
established between $f$ and $F$ provided the parameter $v$ which 
controls the value of $\Phi$ at the UV brane is also small. In 
particular, if $\epsilon$ is of order a tenth and the other parameters 
are of order their natural strong coupling values, then $v$ of order a 
tenth suffices to generate the hierarchy between the Planck and weak 
scales.

After $\varphi$ acquires a VEV, we parametrize the physical 
radion $\widetilde{\varphi}$ that corresponds to the fluctuations over 
the minimum as $\varphi=f\exp\left[\widetilde{\varphi}/f\right]\approx 
f+ \widetilde{\varphi}$. The mass of the radion $\widetilde{\varphi}$ is 
given to leading order in $\epsilon$ as
 \begin{align}
m_{\varphi}^2
&=-\frac{\epsilon \tau}{6}\frac{k^3}{M_5^3}\Big(ke^{-kr_c\pi}\Big)^2 \; .
\label{eq:rad_mass_m}
 \end{align}
 The warped down curvature scale $ke^{-kr_c\pi}\equiv\widetilde{k}$ 
appears naturally in this expression, and is identified with the KK 
scale $m_{KK}$ in the $4D$ theory. Including higher order terms 
in $V(\varphi)$ does not alter the results. At the minimum, the value of 
$T_h$ is tuned to set the $4D$ cosmological constant to zero. Therefore, 
in the presence of a GW mechanism, only one fine-tuning is needed, which 
is the usual tuning to solve the cosmological constant problem.

We see from this analysis that $m_\varphi^2\sim|\epsilon|$, which is 
consistent with previous studies~\cite{Konstandin:2010cd,Eshel:2011wz}. 
Either sign of $\epsilon$ can give rise to a positive mass squared for 
the radion. In the case of $\epsilon>0$ we require $\tau<0$ and $\alpha 
v>0$, while for $\epsilon<0$ these conditions are reversed. (The 
condition on the sign of $\alpha v$ arises from requiring the 
consistency of the minimization condition.) From the holographic 
perspective, the case $\epsilon < 0$ when the solution for $\Phi$ grows 
in the IR is particularly interesting, and is our primary focus.  The 
parametric dependence of the dilaton mass on $\epsilon$ is robust and 
independent of the form of the potential for $\Phi$ on the visible 
brane. Since $\epsilon$ is small this would seem to suggest that the 
presence of a light dilaton is natural in this class of theories. 
However, this result assumed that the cubic self-interaction terms in 
the bulk potential for $\Phi$ could be neglected in favor of the mass 
terms. For this assumption to be self-consistent we require that the 
condition
 \begin{equation}
\frac{1}{3!}\,\eta\,\Phi^3\, \ll \, \frac12m^2\Phi^2
 \end{equation}
 be satisfied at all points in the bulk. Using the solution for $\Phi$ 
and noting that it has the largest magnitude at $\theta=\pi$, this translates 
to
 \begin{equation}
\label{masstermdominant}
\left|
\left(\frac{\eta}{8\sqrt{k}}\right)
\left(\tau+\frac{\alpha^2}{2}\right)
\right|
\ll
3\left|\epsilon\alpha\right|.
 \end{equation}
 For natural values of the parameters $\alpha$, $\tau$, and $\eta$, this 
condition is not expected to be satisfied. Therefore, in determining 
whether a light radion is present in the spectrum, it is in general 
necessary to consider the bulk self-interaction terms.

However, there does exist a class of theories in which the condition 
Eq.~\eqref{masstermdominant} can naturally be satisfied. These are 
theories where the GW scalar is the pNGB of an approximate global 
symmetry that is spontaneously broken. Our results show that in such 
theories the radion can naturally be light, its mass scaling as 
$\sqrt{\epsilon}$, in the limit that back-reaction is small. Although 
in our analysis we only worked to linear order in the small parameter 
$v$, we have explicitly verified that this conclusion is not altered 
when higher powers of $v$ are included.

It is interesting to ask whether the conclusion that the radion is light 
continues to hold when the detuning of the IR brane tension away from 
the RS value is large, so that back-reaction cannot be neglected. 
Although our analysis, strictly speaking, is not valid in this regime, 
the fact that the radion mass continues to scale as $\sqrt{\epsilon}$ 
even as the detuning is taken towards its natural value suggests that this 
result does continue to hold. Although this is not conclusive, evidence 
from the CFT side of the correspondence also indicates that this is the 
case~\cite{Chacko:2012sy}.

\subsubsection*{Massive GW Scalar with a Cubic Interaction in Bulk}

 Next we consider the effect of including self-interaction terms in the 
bulk for $\Phi$. In general, the bulk potential is expected to include 
all powers of $\Phi$, as in Eq.~\eqref{eq:bulk_pot}. However, we 
limit our analysis to the case of a cubic self-interaction term,
 \begin{eqnarray}
V_b(\Phi)=\frac12m^2\Phi^2+\frac{1}{3!}\eta\,\Phi^3 \; .
\label{eq:cubic-potential}
 \end{eqnarray} 
 This restriction is not expected to affect our conclusions because, in 
the limit that the parameters are such that the back-reaction is small, the trilinear term is expected to 
dominate. 
The reason is that small back-reaction implies that the VEV of $\Phi$ is small in units of AdS curvature, so even though all the coefficients of the interaction terms are of order their strong coupling values, the potential will be dominated by the interaction involving the lowest power of $\Phi$, the trilinear term.
 
 With the bulk potential given by Eq.~\eqref{eq:cubic-potential}, the OR equation now becomes
 \begin{eqnarray}
\frac{d\:\log\Phi}{d\:(kr_c\theta)}= 
-\epsilon -\frac{\eta}{8\sqrt{k}}\left(\frac{\Phi}{k^{3/2}}\right) \; .
\label{eq:cubic-potential-il-eqn}
 \end{eqnarray}
 We focus on the case when $\eta/8\sqrt{k}$ is close to its natural 
strong coupling value. We further work in the limit that 
$\big(\eta/8\sqrt{k}\big)\big(\Phi/k^{3/2}\big)$ is large enough that 
$\epsilon$ can be neglected in Eq.~\eqref{eq:cubic-potential-il-eqn}. Although this latter 
condition is not necessarily satisfied at all points in the bulk, it is 
expected to be satisfied in the neighborhood of the IR brane. The reason 
is that in the theories of interest, while $\epsilon$ is required to be 
small in order to generate a large hierarchy, there is no corresponding 
condition on the cubic self-coupling. Since the radion wave function is 
localized near the IR brane, neglecting $\epsilon$ does not 
significantly affect the radion dynamics.

In the limit where the cubic dominates the potential, the OR solution is given by
 \begin{eqnarray} 
\Phi_{OR}(\theta)= \frac{k^{3/2}v}{1+\xi kr_c\theta} \; .
\label{eq:cubic-OR} 
 \end{eqnarray} 
 Here $\xi = \eta v/8\sqrt{k}$ and we have imposed the boundary 
condition $\Phi(0)=k^{3/2}v$. In order for our boundary layer analysis 
to be self-consistent, we require that both $\alpha$ and $v$ 
to lie below their natural values by a factor that could be as small as 
a few. However, we have already chosen $v$ to be small in order to 
generate a large hierarchy. We have also set $\alpha$ somewhat smaller 
than its natural value so that gravitational back-reaction can be 
neglected. Hence our boundary layer analysis is valid without additional 
restrictions. Combining the OR and the BR solutions, we obtain the 
complete solution for $\Phi$ as
 \begin{eqnarray} 
\Phi(\theta)=-\frac{k^{3/2}\alpha}{4}e^{4kr_c(\theta-\pi)}
+\frac{k^{3/2}v}{1+\xi kr_c\theta} \; .
\label{eq:cubic-potential-phi-complete} 
 \end{eqnarray} 
 We can now compute the potential for the radion to leading order in $v$:
 \begin{align} 
V(\varphi)=k^4\left(\frac\varphi F\right)^4
\left[
\tau+\frac{1}{1-\xi\log(\varphi/F)}\left( 2\alpha v +\frac{\alpha^2 \xi}{8}\right) \right] \; .
\label{eq:rad-potential-cubic} 
 \end{align} 
 We can define $w=2\alpha v +\alpha^2 \xi/8$ and note that $w$ is small because it is linear in $v$. The potential is minimized when the 
condition
 \begin{eqnarray} 
\tau+\frac{w}{1-\xi\log(f/F)}=0
\label{eq:minimize_cubic} 
 \end{eqnarray}
is satisfied. We can solve for $\langle \varphi \rangle = f$,
 \begin{eqnarray}
\left[\frac{f}{F}\right]^\xi\equiv e^{-\xi kr_c\pi}
&=&e^{1+w/\tau}+\mathcal{O}(\xi) \; .
 \end{eqnarray} 
 This is more transparent when written in terms of the radius of the 
extra dimension,
 \begin{eqnarray}
k \pi r_c  \approx -\frac{w+ \tau }{\xi \tau} 
           \approx -\frac{1}{\xi} \;.
\label{eq:length}
 \end{eqnarray} 
 To obtain a large warp factor $k\pi r_c$ must be larger than 1. This 
condition is satisfied if $v$ is of order $10^{-2}$, which justifies the last approximation in Eq.~\eqref{eq:length}. From 
Eqs.~\eqref{eq:rad-potential-cubic} and~\eqref{eq:minimize_cubic}, we 
can compute the mass squared of the radion
 \begin{eqnarray} 
m_\varphi^2 &=&\frac{\tau^2\xi}{6w }\, \frac{k^3}{M_5^3}
\left(ke^{-kr_c\pi}\right)^2 .
\label{eq:rad_mass_cubic} 
 \end{eqnarray}
 Details of the calculation are given in appendix~\ref{app:b}.
 
Putting in the natural values of the parameters given in 
appendix~\ref{app:NDA}, we find that the radion mass is parametrically 
of order the KK scale, so that it is no longer light. Although our 
analysis neglected back-reaction, we do not expect that including this 
effect would alter our conclusions. If $\eta$, (as well as $\epsilon$), 
was tuned to be small, then Eq.~\eqref{eq:rad_mass_cubic} would predict 
a light radion. In that case, however, the quartic and higher order 
terms in the $\Phi$ bulk potential would give large contributions to the 
radion mass. If, however, all the terms in the $\Phi$ bulk potential are 
small in units of curvature, the radion will be light. This could be 
realized naturally, if, for example, the GW field corresponds to a 
pseudo-Goldstone boson.

We also see from Eq.~\eqref{eq:rad_mass_cubic} that the mass of the 
radion scales as $\tau$. Therefore a light radion can arise if $\tau$ 
lies below its natural strong coupling value. Small values of $\tau$ are 
associated with tuning, since this condition is not expected to be 
satisfied in general. However, the tuning of the radion mass is 
mild, scaling as $\tau$. Therefore a radion that lies a factor of 5 
below the KK scale is only tuned at the level of 1 in 5 (20\%).

\subsection{Mixing with Other Scalars}
\label{sec:mixing}

Along with the radion, the other scalars in the low energy spectrum are 
the first few KK modes $\{\phi_n\}$ of the scalar $\Phi$ which in 
general mix with the physical radion $\widetilde{\varphi}$. Expanding 
the $\Phi$ action to linear order in $\widetilde{\varphi}$ and $\phi_n$, 
all terms without a spacetime derivative vanish because of the classical 
$\Phi$ equations of motion~\cite{Rattazzi:2000hs}. This implies there is only kinetic mixing 
between $\widetilde{\varphi}$ and $\phi_n$. The mixing coefficient 
$\kappa_n$ affects the physical radion and KK modes as
 \begin{align}
\widetilde{\varphi}
&\rightarrow\widetilde{\varphi}-\kappa_n\frac{m_n^2}{m_n^2-m_\varphi^2}\phi_n
\qquad\qquad
\phi_n
\rightarrow\phi_n-\kappa_n\frac{m_\varphi^2}{m_\varphi^2-m_n^2}\widetilde{\varphi}\; ,
\label{eq:3.6}
 \end{align}
 where $m_n$ is the mass of the KK mode and $\widetilde{\varphi}$ is the 
radion field. We find that $\kappa_n \sim m_\varphi^2/m_{n}^2$ for both types of bulk $\Phi$ potentials considered in this section. 

\section{Radion Couplings}
\label{sec:rad-coupling}

In this section we calculate the corrections to the radion couplings to 
SM fields that arise as a consequence of the GW stabilization mechanism. 
In the absence of the GW scalar, the radion being a massless scalar 
gravitational mode couples to the trace of the $5D$ energy momentum 
tensor~\cite{Goldberger:1999un}. Its couplings are therefore completely 
fixed. This has been used to determine the couplings of the radion, both 
in the case when the SM fields are confined to the IR 
brane~\cite{Goldberger:1999uk,Goldberger:1999un,Csaki:1999mp, 
Giudice:2000av, Csaki:2000zn} and the case when they reside in the bulk 
of the space~\cite{Csaki:2007ns, Rizzo:2002pq}. In this paper we will 
consider SM fields confined to a brane, and leave the case of bulk fields 
for future work.

Once the GW scalar $\Phi$ is added to the 
theory, in general we expect interactions that couple 
$\Phi$ to the SM fields. Since the VEV of $\Phi$ depends on the brane 
spacing, and therefore on the background radion field, this effect 
contributes to the coupling of radion to SM fields. To understand this in detail, consider an arbitrary term in the 
Lagrangian that depends only on the SM fields,
 \begin{align}
\mathcal{L} \supseteq \sqrt{G}
\;f(\Psi_i, \mathcal{A}^\mu_i,\mathcal{H}_i) \;.
\label{eq:4.01}
 \end{align}
 Here $f(\Psi_i,\mathcal{A}^\mu_i,\mathcal{H}_i)$ is a function of the SM fermions $\Psi_i$, the SM gauge fields $\mathcal{A}^\mu_i$ and the Higgs field $\mathcal{H}$. Then, one can also write the following interaction term in the Lagrangian involving the GW field $\Phi$,
 \begin{align}
\mathcal{L} \supseteq  \alpha_{\rm int}\:\sqrt{G}
\;f(\Psi_i, \mathcal{A}^\mu_i,\mathcal{H}_i)\:k^{-3/2}\Phi(x,\theta)\; ,
\label{eq:4.1}
 \end{align}
 where $\alpha_{\rm int}$ is a dimensionless coupling constant. Terms 
containing higher powers of $\Phi$ in the interaction do not change our 
conclusions as we work to leading order in $v$. In theories with a 
holographic dual, we expect $\alpha_{\rm int}$ to be of order its natural 
strong coupling value (see appendix~\ref{app:NDA}).


In order to determine the radion couplings, we allow the GW field to 
fluctuate and KK expand the fluctuations over the classical value. This 
amounts to replacing
 \begin{eqnarray}
\Phi(\theta)\rightarrow\Phi(\theta)
+\sum_n f_n(\theta)\phi_n(x)
\label{eq:}
 \end{eqnarray} 
 in Eq.~\eqref{eq:4.1}, where $\phi_n$ are the KK modes of the GW field 
$\Phi$. The radion couplings to the light fields contained in $f(\Psi_i, 
\mathcal{A}^\mu_i,\mathcal{H}_i)$ get contributions from two 
sources. Firstly, in the parametrization that we consider, the 
background value $\Phi(\theta)$ is a function of the background radion 
field $\varphi(x)$ after $r_c$ is made dynamical. Expanding 
$\Phi(\theta)$ about the radion VEV to linear order generates coupling 
to the physical radion $\widetilde{\varphi}\approx\varphi-f$ as 
 \begin{eqnarray} 
\Phi(\theta)\rightarrow\Phi(\varphi,\theta)=\Phi(f,\theta) 
+\widetilde{\varphi}\:\partial_\varphi\Phi(f,\theta)+... \;\;\; .
\label{eq:rad-fluc} 
 \end{eqnarray} 
 Secondly, the KK modes $\phi_n$ in general have a kinetic mixing 
with the radion, and thereby generate a radion coupling to the SM 
fields. From section~\ref{sec:mixing}, it follows that this effect scales as $m_\varphi^4/m_{KK}^4$ for a light radion. As we will see below, this effect is subleading and can be neglected.

We can schematically understand how the corrections to the radion 
couplings scale. For simplicity, we focus on the case where the SM 
fields live on the visible brane. When the bulk potential for $\Phi$ 
is dominated by the mass term, we find that the leading source of 
modification to the radion coupling, using Eqs.~\eqref{eq:rad-fluc} 
and~\eqref{eq:massive-phi-complete}, scales as 
$\partial_\varphi\Phi|_{\theta=\pi}\sim\epsilon\:v\: e^{-\epsilon 
kr_c\pi}$. From the minimization condition Eq.~\eqref{eq:2.7} and the 
expression for the mass of the radion Eq.~\eqref{eq:rad_mass_m}, we obtain
 \begin{eqnarray}
\epsilon\:v\:e^{-\epsilon kr_c\pi}\sim {\epsilon\tau}
\sim\frac{m_\varphi^2}{\:m_{KK}^2} \; .
\label{eq:minimiz-mass}
 \end{eqnarray}
 We see that the correction scales as $\sim m_\varphi^2/m_{KK}^2$ and is small for a light radion.

In the case when the the bulk potential for $\Phi$ is dominated by the 
cubic self interaction term, using Eqs.~\eqref{eq:rad-fluc} 
and~\eqref{eq:cubic-potential-phi-complete}, we find that the leading source of 
corrections to the radion coupling now scales as 
 \begin{equation}
\partial_\varphi\Phi|_{\theta=\pi} \sim v\xi/(1-\xi\log f/F)^2 \; . 
 \end{equation}
 From the expressions for the minimization 
condition Eq.~\eqref{eq:minimize_cubic} and the mass of the 
radion Eq.~\eqref{eq:rad_mass_cubic}, we find that
 \begin{eqnarray}
\frac{v\xi}{\left(1-\xi\log(f/F)\right)^2}\sim \tau^2
\sim\frac{m_\varphi^2}{\:m_{KK}^2} \; ,
\label{eq:}
 \end{eqnarray} 
 so that the corrections again scale as $\sim m_\varphi^2/m_{KK}^2$. It 
follows from this analysis that the form and magnitude of the leading 
corrections to the radion couplings does not depend on the details of 
the GW mechanism that generates the radion mass.

We now calculate the 
corrections to the radion couplings to SM fields in detail, focusing on 
the radion couplings to massive and massless gauge bosons, and to 
fermions.  We will consider the 
case where the Higgs is the pNGB of an 
approximate global symmetry~\cite{Georgi:1974yw,Georgi:1975tz,Kaplan:1983fs,Kaplan:1983sm,Georgi:1984af}. In this scenario, mixing between the Higgs 
and the radion, which can otherwise be significant, is 
suppressed~\cite{Giudice:2000av}, and therefore the Higgs can be 
replaced by its VEV in the action. This also allows the results from 
this section to be directly carried over to Higgsless models~\cite{Csaki:2003zu}.

\subsection{Massive Gauge Bosons}

We begin by considering the radion couplings to the massive gauge bosons 
of the SM, the $W^{\pm}$ and the $Z$. In our discussion, we shall focus 
exclusively on the $W^{\pm}$, the generalization to the $Z$ being 
straightforward. In the limit when the effects of brane stabilization 
are neglected, the relevant terms in the action for the $W$ bosons take 
the form
 \begin{eqnarray}
\mathcal{S}
&=&
\int d^4x\;d\theta\;\delta(\theta-\pi)\;\sqrt{-G_v}
\left[
-\frac{1}{4g^2} G_v^{\mu\rho}G_v^{\nu\sigma} W_{\mu\nu}W_{\rho\sigma}
+G_v^{\mu\nu} D_{\mu}H^\dagger D_{\nu}H 
\right] \; ,
\label{eq:Gauge-Lag}
\end{eqnarray} 
 where $G_v$ is the induced metric on the visible brane and $g$ is the gauge coupling. Here we use the convention where the gauge covariant derivative is given by $D_\mu=\partial_\mu +A_\mu$, and the gauge coupling appears in the kinetic term of the gauge fields. 
 When $H$ is 
replaced by its VEV in Eq.~\eqref{eq:Gauge-Lag}, the $W$ bosons acquire 
a mass. Expanding out the components of the metric in terms of the 
radion and graviton, we find that the radion couples to the $W^{\pm}$ as
 \begin{eqnarray}
\frac{2m_W^2}{g^2}\:\frac{\widetilde{\varphi}}{f}\:W_\mu^+ W^{\mu-} \;,
\label{eq:mass-gauge-radion}
 \end{eqnarray}
 where the index on $W$ is now raised by the $4D$ Minkowski metric of flat 
space. Just as for the SM Higgs, at tree level the coupling of the 
radion is proportional to the mass of the field it is coupling to.
 
In the presence of the GW field, the gauge kinetic term on the brane is 
modified to
 \begin{align} 
\mathcal{L} \supseteq \delta(\theta - \pi)\:\sqrt{-G_v} 
\left[ 
-\frac{1}{4 \hat{g}^2} G_v^{\mu\rho}G_v^{\nu\sigma} W_{\mu\nu}W_{\rho\sigma}
\left( 1 + \alpha_W \frac{\Phi}{k^{3/2}} \right) 
\right] \; .
\label{eq:phi-gauge}
 \end{align}
 Here $\hat{g}$ represents the gauge coupling in the absence of the GW 
stabilization mechanism, and $\alpha_W$ is a dimensionless number. We 
take $\alpha_W$ to be its natural value but continue to work in a regime 
where $v$ is small and work to leading order in $v$. This turns out to 
be equivalent to working to linear order in $\alpha_W$. The physical 
gauge coupling is now given by 
 \begin{eqnarray} 
 \frac{1}{4g^2} = \frac{1}{4 \hat{g}^2} 
\Big( 1 + \alpha_W \frac{\Phi(\pi)}{k^{3/2}} \Big) \; .
\label{eq:gauge_correction} 
 \end{eqnarray} 
 To incorporate radion fluctuations about the VEV of $\Phi$, we let
 \begin{eqnarray} 
\Phi(\pi)\rightarrow \Phi(\pi) 
\left(1+\frac{\partial_\varphi\Phi(\pi)}{\Phi(\pi)}\:\widetilde{\varphi}\right) \; .
 \end{eqnarray} 
 Using the classical solution for $\Phi$, the interaction term leads to 
a correction to the radion coupling. We consider first the case when the 
bulk potential for $\Phi$ is dominated by the mass term and the 
self-interactions can be neglected. In this limit, the radion coupling 
is given by
 \begin{align} 
\bar{c}_W \left( \epsilon \: v\:e^{-\epsilon kr_c\pi} \right) 
\frac{1}{4g^2} \frac{\widetilde{\varphi}}{f}\:W_{\mu\nu} W^{\mu \nu} \; .
\label{eq:gauge-radion} 
 \end{align} 
 Here $\bar{c}_W$ is an $\mathcal{O}(1)$ number, and indices on $W$ are 
raised by the Minkowski metric $\eta^{\mu\nu}$. Using the minimization 
condition and the formula for the mass of the radion, we have 
$v\:\epsilon\:e^{-\epsilon kr_c\pi} \sim \epsilon\:\tau \sim 
m_\varphi^2/m_{KK}^2$. Therefore the correction scales as 
$m_\varphi^2/m_{KK}^2$, and is small for a light radion.\footnote{We note that the operators in Eq.~\eqref{eq:mass-gauge-radion} and Eq.~\eqref{eq:gauge-radion} are different, and that it may be possible to tell their contributions apart using gauge boson polarizations, especially at high energy.}

The GW scalar also couples to the gauge covariant kinetic term for the 
Higgs. These interactions affect the gauge boson masses, and lead to 
corrections to the radion couplings to these fields.  We therefore 
consider the term
 \begin{align}
\mathcal{L} \supseteq \delta(\theta - \pi)\:\sqrt{-G_v}
\left[ 
\left( 1 + \beta_W \frac{\Phi}{k^{3/2}} \right)
G_v^{\mu\nu} D_{\mu}H^\dagger D_{\nu}H  
\right] \; ,
 \end{align} 
 where $\beta_W$ is a dimensionless number. Working in unitary gauge, we 
replace the Higgs field $H$ by its VEV. The physical gauge boson mass is 
now modified to
 \begin{eqnarray}
m_W^2 = 
\hat{m}_W^2
\frac{\Big(1+\beta_W \Phi(\pi)/k^{3/2}\Big)}
{\Big(1+\alpha_W \Phi(\pi)/k^{3/2}\Big)} \; ,
\label{eq:4.12}
 \end{eqnarray}
 where $\hat{m}_W$ is defined as the gauge boson mass in the absence of 
the GW stabilization mechanism. Including the radion fluctuations about 
the VEV of $\Phi$, we obtain for the radion coupling
 \begin{align}
\frac{\widetilde{\varphi}}{f}
\left[
2 + c_W \: (\epsilon \: v e^{-\epsilon kr_c\pi} )
\right]
\frac{m_W^2}{g^2}\:W_\mu^+ W^{\mu-} \; .
\label{eq:4.13}
 \end{align}
 Here $c_W$ is an $\mathcal{O}(1)$ number. We see that this correction is also
suppressed by $m_\varphi^2/m_{KK}^2$, and is small for a light radion.

For the case when the bulk potential for $\Phi$ has a cubic interaction 
and no mass term, we can repeat the steps above to obtain the 
corrections to the radion coupling. The end result however remains the 
same, with the corrections again scaling as $m_\varphi^2/m_{KK}^2$.

\subsection{Massless Gauge Bosons}

Next we consider the case of radion couplings to the massless gauge bosons of the SM, the photon and the gluons, on the IR brane. The kinetic term has the same form as Eq.~\eqref{eq:Gauge-Lag}, but now the coupling to the Higgs is absent.  Expanding out the components of the metric in terms of the $4D$ graviton and radion, we find that the radion does not couple to the massless gauge bosons. However this statement is true only at the classical level. In general, quantum effects generate a coupling of the massless gauge bosons to the radion at one loop. This can be understood as arising from quantum corrections to the trace of the energy momentum tensor; the trace anomaly. These contributions have been calculated in~\cite{Giudice:2000av}, and the resulting coupling is given by
\begin{equation}
\frac{b_<}{32\pi^2}
\frac{\widetilde{\varphi}}{f}
F_{\mu\nu}F^{\mu\nu} \; ,
\label{eq:4.10a}
\end{equation}
where again the indices are raised by $\eta^{\mu\nu}$. Here $b_<$ is proportional to the one-loop $\beta$-function coefficient for the running of the gauge coupling, 
\begin{equation}
\frac{d}{d \; {\rm log} \mu} \frac{1}{g^2} = \frac{b_<}{8 \pi^2} \; .
\label{eq:beta}
\end{equation}
 This formula is valid at the KK scale and must be renormalization 
group (RG) evolved to the radion mass to determine the couplings to the 
photon and gluons for the on-shell radion.

In the presence of the GW stabilization mechanism, there is a coupling between $\Phi$ and the gauge bosons given in Eq.~\eqref{eq:phi-gauge}, which modifies the gauge coupling as in Eq.~\eqref{eq:gauge_correction}.  In the case where the GW scalar has a bulk mass term, a coupling between the radion and gauge bosons of the form of Eq.~\eqref{eq:gauge-radion} is generated by the stabilization dynamics. This coupling is of order $m_\varphi^2/m_{KK}^2$, and, in the case where the bulk potential of $\Phi$ has no mass but a cubic interaction, it is straightforward to verify that the correction is of the same form and again scales as $m_\varphi^2/m_{KK}^2$. Although the corrections arising from stabilization are small, the fact that the leading order effect is loop suppressed implies that when the radion is only moderately lighter than the  Kaluza-Klein scale, the effects of GW stabilization are significant, and may even dominate.

\subsection{Fermions}

We finally consider the case of the radion couplings to brane localized 
SM fermions. For concreteness we will focus on the radion couplings to 
up-type quarks, $Q$ and $U$. The generalization to the other fermions is 
straightforward. In the absence of a stabilization mechanism, the 
relevant part of the action has the form
 \begin{align} 
\mathcal{S} &= \int 
d^4x\;d\theta\;\delta(\theta-\pi)\;\sqrt{-G_v} \left[ \frac{i}{2}e_a^\mu 
\Big( \bar{Q}\:\Gamma^a\overleftrightarrow{\partial_\mu} Q + 
\bar{U}\:\Gamma^a\overleftrightarrow{\partial_\mu} U \Big) 
-y\Big(\bar{Q}HU+\bar{U}H^\dagger Q\Big) \right] \; ,
\label{eq:4.2} 
\end{align} 
where $\overleftrightarrow{\partial} = \overrightarrow{\partial}-\overleftarrow{\partial}$ and $e^\mu_a$ is the vierbein. We replace $H$ by its VEV and expand the components of the metric and vierbein out in terms of the $4D$ graviton and radion. After expanding $\varphi$ about its VEV, canonically normalizing the kinetic terms of the fields $Q$ and $U$, and using the equations of motion for the fermions~\cite{Arzt:1993gz, Georgi:1991ch}, we obtain the coupling of the fermions to the radion as
\begin{eqnarray} 
-m_f\frac{\widetilde{\varphi}}{f}\left(\bar{Q}U+h.c.\right) \; ,
\label{eq:} 
\end{eqnarray} 
showing that the radion couples proportional to mass as expected. 

The GW stabilization mechanism allows the following additional 
interaction terms involving $\Phi$,
 \begin{align} 
\mathcal{L}_{\rm int} 
=\delta(\theta-\pi)\sqrt{-G_v} \: \frac{\Phi}{k^{3/2}} 
\left[ \frac{i}{2}e_a^\mu 
\Big( \alpha_q\bar{Q}\:\Gamma^a\overleftrightarrow{\partial_\mu} Q + 
\alpha_u\bar{U}\:\Gamma^a\overleftrightarrow{\partial_\mu} U \Big) 
-\beta\:y\Big(\bar{Q}HU+\bar{U}H^\dagger Q\Big) \right] \; ,
\label{eq:} 
 \end{align} 
 where $\alpha_q,\alpha_u$ and $\beta$ are dimensionless numbers. After 
$\Phi$ gets a VEV, the kinetic and mass terms receive corrections. The 
fluctuations of $\Phi$ about its VEV give rise to corrections to the 
radion couplings. We first focus on the case when the bulk potential for 
$\Phi$ consists only of a mass term. After making the kinetic terms 
canonical, and using the equations of motion for the fermions, the 
coupling to $\widetilde{\varphi}$ is determined to be of the form
 \begin{align} 
-m_f\:\frac{\widetilde{\varphi}}{f}\Big(\bar{Q}U+h.c.\Big) 
\left[ 1+c_\psi\:\epsilon\:v\:e^{-\epsilon kr_c\pi} \right] \; . 
\label{eq:4.7} 
 \end{align} 
 Here $c_\psi$ is an $\mathcal{O}(1)$ number. We see that the 
corrections are suppressed by $\sim m_\varphi^2/m_{KK}^2$, and are 
therefore small for a light radion. In the case of a cubic bulk 
potential for $\Phi$, the corrections are of the same form and 
also suppressed by $\sim m_\varphi^2/m_{KK}^2$.

\section{The Dual Picture}
\label{sec:dual-pic}

In this section, we apply the rules of the AdS/CFT dictionary
to compare 
the results obtained in this paper for the mass and the couplings of the 
radion to those in the literature for its holographic dual, the 
dilaton~\cite{Chacko:2012sy,Bellazzini:2012vz}, and we find good agreement. In the 
process, we are also able to arrive at a holographic interpretation of 
our results for the radion. In particular, we will establish that the 
bulk self-interactions for $\Phi$ that seemed somewhat ad-hoc and 
non-minimal from the higher dimensional viewpoint, are well motivated in 
the dual picture, and are in fact necessary to obtain agreement. Our 
approach in this section will be to review the 
calculations for the mass and the couplings of the dilaton in a strongly 
coupled CFT and to compare against the corresponding results for 
the radion from earlier sections of this work.

In the AdS/CFT dictionary, the coordinate corresponding to the fifth 
dimension of AdS space is associated with the renormalization scale 
$\mu$ in the dual theory. To make this more precise, consider making a 
change of coordinates in AdS space from $\theta$ to $z$, where $z$ is 
defined as
 \begin{equation}
z=\frac{e^{kr_c\theta}}{k}\; .
 \end{equation}
 Then the renormalization scale $\mu$ in the dual CFT is given by $\mu 
\sim 1/z$. Therefore, the RS set-up with two branes is dual to a strongly 
coupled theory that is well approximated by a CFT in the energy regime 
between the two branes. The hidden brane corresponds to the UV cut-off 
of the theory. The boundary conditions on the bulk fields at this 
boundary determine the coefficients of the deformation of the CFT in the 
UV, in the dual picture. The visible brane ends the AdS space in the IR, 
signaling the breakdown of the CFT. Various checks can be performed that 
are suggestive of a spontaneous breakdown. In particular, the trace of 
the energy momentum tensor is unchanged by the presence of the visible 
brane. Furthermore, the two point function of the dilatation current has 
a massless pole of the appropriate strength as dictated by Goldstone's 
theorem~\cite{Rattazzi:2000hs}. The scalar excitation associated with 
this massless pole, the radion, is identified with the NGB of broken 
scale invariance, the dilaton.

The AdS geometry is stabilized by adding a GW scalar $\Phi$ to the 
theory. In the dual picture, this corresponds to deforming the CFT by a 
primary scalar operator $\mathcal{O}$. The boundary condition for $\Phi$ 
on the UV brane is related to the strength of the deformation. In what 
follows, we compare the potentials for the radion and the dilaton on the 
two sides of the duality. We first focus on the original model where 
there is no stabilization mechanism before considering the scenario when 
the GW scalar is present. In both cases, we obtain good agreement. We 
make use of the ideas of holographic 
renormalization~\cite{Porrati:1999ew,Balasubramanian:1999jd,Verlinde:1999fy,de-Boer:1999xf,Verlinde:1999xm,Balasubramanian:2000wv,de-Haro:2000xn} 
when identifying parameters on the two sides of the correspondence.

\subsection{Dynamics in the Absence of a Stabilization Mechanism}

The dilaton is the NGB associated with the spontaneous breaking of 
conformal symmetry. Under the scale transformation $x^{\mu} \rightarrow 
x'^{\mu} = e^{- \omega} x^{\mu}$, the dilaton undergoes a shift, 
$\sigma(x) \rightarrow \sigma'(x') = \sigma(x) + \omega f$. Here $f$ is 
the scale associated with the breaking of conformal symmetry. For the 
purpose of writing interactions of the dilaton it is convenient to 
define the object
 \begin{equation}
\chi(x) = f e^{\sigma(x)/f}\;,
 \end{equation}
 which transforms linearly under scale transformations. Specifically,
under the scale transformation $x^{\mu} \rightarrow x'^{\mu} =
e^{-\omega} x^{\mu} $, $\chi(x)$ transforms as a conformal compensator
 \begin{equation}
\chi(x) \rightarrow \chi'(x') = e^{\omega} \; \chi(x) \; .
 \end{equation} This symmetry forbids a mass term for $\chi$. However, 
unlike the NGBs associated with a spontaneously broken internal symmetry, 
the potential for $\chi$ admits a quartic term
 \begin{eqnarray}
V(\chi)=\kappa_0\:\frac{\chi^4}{4!} \; .
\label{eq:dilaton-potential}
 \end{eqnarray}
 Unless $\kappa_0$ is fine-tuned to zero, this potential does not admit a 
stable minimum away from the origin.

We now consider the radion potential, which is very similar. Before stabilization, $\varphi$ is massless in AdS and corresponds to 
the freedom to alter the distance between the two branes. The potential 
for $\varphi$ in the absence of stabilization is given by (see 
appendix~\ref{app:a})
 \begin{eqnarray}
V(\varphi)= \frac{k^4}{F^4}\tau\:\varphi^4 \;.
\label{eq:}
 \end{eqnarray}
 We see that the forms of the potential for the dilaton and the radion 
are identical.
On the AdS side,
$\tau=(T_v-\Lambda_b/k)/k^4$, and we have
\begin{equation}
\tau \;\Leftrightarrow \; \kappa_0 \;,
\end{equation}
where we use the double arrow to denote that these two quantities are related by the duality. In the 
original RS solution, the IR brane tension $T_v$ is chosen to 
set $\tau=0$. In the dual theory, this is equivalent to fine-tuning 
$\kappa_0$ to vanish.

\subsection{Dynamics in the Presence of a Stabilization Mechanism}

In the dual description, the addition of the GW scalar corresponds to 
deforming the CFT by a primary scalar operator $\mathcal{O}$ that has
scaling dimension $\Delta$ in the far UV,
 \begin{eqnarray}
\mathcal{L}_{CFT}
\rightarrow
\mathcal{L}_{CFT}+\hat{\lambda}(\mu)\:\mu^{4-\Delta}\:\mathcal{O}(x) \; .
\label{eq:}
 \end{eqnarray}
 Here $\hat{\lambda}(\mu)$ is the renormalized coupling constant. 
Defined in this way it is dimensionless. We choose to normalize the 
operator $\mathcal{O}$ such that $\hat{\lambda}$ of order one 
corresponds to conformal symmetry violation becoming strong, so that it 
can no longer be treated as a perturbation. If the operator 
$\mathcal{O}$ is relevant ($\Delta<4$), the deformation at the UV scale 
$\Lambda_{\rm UV}$ grows in the IR leading to the breaking of the CFT at 
the IR scale $f$. Below the breaking scale the spectrum contains the 
dilaton $\chi$, which is the Goldstone boson of spontaneously broken 
conformal symmetry. As a consequence of the explicit conformal symmetry 
violation associated with the operator $\mathcal{O}$, the potential for 
$\chi$ gets modified from Eq.~\eqref{eq:dilaton-potential}. This results 
in $\chi$ acquiring a mass. The potential for the dilaton $\chi$ in the 
presence of this explicit breaking of CFT can be determined from a 
spurion analysis. In the presence of the deformation, the coefficient 
$\hat{\lambda}$ satisfies a renormalization group (RG) equation above 
the symmetry breaking scale
 \begin{eqnarray}
\frac{d \log \hat{\lambda}}{d \log \mu} = -g(\hat{\lambda}) \;,
\label{eq:g-def}
 \end{eqnarray} where $g(\hat{\lambda})$ is a polynomial in 
$\hat{\lambda}$ that can be parametrized as
 \begin{eqnarray}
g(\hat{\lambda})=(4-\Delta)+c_1\hat{\lambda}+c_2\hat{\lambda}^2+... \;\;\; .
\label{poly}
 \end{eqnarray}
With our normalization of the operator $\mathcal{O}$, the coefficients
$c_i$ are expected to be of order one. 
 
 The cases of interest for electroweak symmetry breaking correspond to 
those where there is a large hierarchy between the cutoff scale 
$\Lambda_{\rm UV}$ and the scale $f$ where conformal symmetry is broken.  
A large hierarchy can be generated if the deformation $\hat{\lambda}$ is 
exponentially small in the UV. This is technically natural if the 
operator $\mathcal{O}$ breaks a symmetry of the CFT, and is therefore 
protected. The AdS/CFT correspondence relates this scenario to the case 
when the value of $\Phi$ at the UV brane is exponentially small. 
Alternatively, a large hierarchy can be generated if the operator 
$\mathcal{O}$ is close to marginal in the UV, so that $|4-\Delta| \ll 
1$. Duality relates this scenario to the case when the bulk mass term 
for the GW scalar is small in units of the curvature. It is this latter 
scenario which we will focus on.

We wish to determine the form of the potential for the dilaton in the 
low energy effective theory. To do this, we note that the UV theory 
possesses a spurious scale invariance if $\hat{\lambda}$ is assigned a 
(spurious) scaling dimension $g(\hat{\lambda})$. The potential for the 
dilaton in the low energy theory must respect this spurious symmetry. 
This is a severe restriction on the possible terms that can appear. 

For the purposes of obtaining the low energy effective theory it is 
useful to construct an object $\overline{\Omega}(\hat{\lambda}, 
\chi/\mu)$ that is invariant under both infinitesimal scale and RG 
transformations,
 \begin{eqnarray}
\overline{\Omega}(\hat{\lambda}, \chi/\mu) = 
\hat{\lambda}\left( 1 - g(\hat{\lambda})\log(\chi/\mu) \right) \; .
\label{eq:}
 \end{eqnarray}
 For values of $\mu$ close to the symmetry breaking scale $f$ and for 
small $g(\hat{\lambda})$, this invariant combination can be approximated 
as
 \begin{eqnarray}
\overline{\Omega}(\hat{\lambda}, \chi/\mu) \approx
\hat{\lambda}\left(\frac{\chi}{f}\right)^{-g(\hat{\lambda})} \; ,
\label{eq:}
 \end{eqnarray} 
 where $\hat{\lambda}$ in this expression is understood to be evaluated 
at $\mu = f$. Then by requiring invariance under spurious scale 
transformations we can obtain the potential for the dilaton. To leading 
order in $\hat{\lambda}$ it is given by~\cite{Chacko:2012sy}
 \begin{eqnarray}
V(\chi)= \frac{\chi^4}{4!}
\left[
\kappa_0
-\kappa_1\: \overline{\Omega}(\hat{\lambda}, \chi/\mu)
\right] \approx  
 \frac{\chi^4}{4!}
\left[
\kappa_0
-\kappa_1\:\hat{\lambda}\left(\frac{\chi}{f}\right)^{-g(\hat{\lambda})}
\right]
\; .
\label{eq:potential-dilaton}
 \end{eqnarray} 
 Here $\kappa_0$ and $\kappa_1$ are parameters in the low energy theory. The 
potential now admits a stable minimum at $\left<\chi\right>=f$ without 
fine-tuning $\kappa_0$ to zero. For $g(\hat{\lambda}) \ll 1$, the 
minimization condition takes the form,
 \begin{eqnarray}
\kappa_0=\kappa_1\:\hat{\lambda} \; .
\label{mincondn}
 \end{eqnarray} 
 In this equation, $\hat{\lambda}$ is evaluated at the breaking scale 
$f$. This implicitly determines the breaking scale. By minimizing this 
potential, the mass of the dilaton $\sigma$ can be obtained as
 \begin{eqnarray}
m_\sigma^2= \frac{4\kappa_0}{4!}\:g(\hat{\lambda})f^2 \; .
\label{eq:}
 \end{eqnarray}
 It follows that the mass of the dilaton depends on $\kappa_0$ and 
$g(\hat{\lambda})$, which are related to each other by Eqs.~\eqref{poly} and~\eqref{mincondn}.  We focus on the 
case when $\kappa_0$ lies below its natural strong coupling 
value by a factor which could be as small as order a few. However the 
other parameters including $\kappa_1$ and the $c_i$ are assumed to be of 
order their natural values. In this limit, $\hat{\lambda}$ at the 
minimum is also expected to lie below its strong coupling value, and by 
roughly the same factor. For sufficiently small $\kappa_0$, 
$g(\hat{\lambda})$ at the minimum is dominated by the constant term in Eq.~\eqref{poly} so so that it scales as $g(\hat{\lambda})\sim(4-\Delta) 
\ll 1$. In this limit the mass squared of the dilaton scales as 
$m_\sigma^2\sim\kappa_0(4-\Delta)$.

However, since in the theories of interest we have $4 - \Delta \ll 1$, 
for larger values of $\kappa_0$ we expect that close to the breaking 
scale $\mu = f$ the term linear in $\hat{\lambda}$ in Eq.~\eqref{poly} 
will dominate over the constant term. In this limit $g(\hat{\lambda})$ 
at the minimum scales as $g(\hat{\lambda}) \sim \hat{\lambda} \sim 
\kappa_0$. We then have $g(\hat{\lambda}) \sim \kappa_0$, so that in 
this limit the mass squared scales as $m_\sigma^2 \sim \kappa_0^2$. 
Hence we see that, depending on the extent to which $\kappa_0$ lies 
below its natural strong coupling value, the dilaton mass may scale 
either as $\sqrt{\kappa_0(4-\Delta)}$ or as $\kappa_0$. Recall that 
duality relates $\kappa_0$ to $\tau$, the detuning of the IR brane 
tension. We therefore see that the results for the scaling of the 
radion mass with the IR brane tension are indeed reproduced on the CFT 
side of the correspondence.

Following this quick review we are ready to draw more detailed parallels 
to the AdS side of the duality in the presence of stabilization. Recall 
that the classical equation satisfied by $\Phi$ has two independent 
solutions, which in the limit that $kr_c \gg 1$, are well approximated 
by the OR and BR solutions defined in section~\ref{sec:rad-mass}. The OR 
differential equation is simply first order, and is a good approximation 
to the classical equation that is satisfied by $\Phi$ everywhere except 
in the boundary layer region close to the IR brane. 
In the absence of an IR brane, the AdS/CFT dictionary relates the value of $\Phi$ at the UV brane to the coefficient of the deformation of the CFT, $\hat{\lambda}$, evaluated at the renormalization scale $\mu = k$~\cite{Gubser:1998bc,Klebanov:1999tb,Rattazzi:2000hs}.
We denote this correspondence by
 \begin{equation}
\frac{\Phi(\theta=0)}{k^{3/2}} \;\Leftrightarrow \;
\hat{\lambda}(\Lambda_{\rm UV})\;.
\label{eq:lambda_corres}
 \end{equation}
 The ideas of holographic renormalization extend this identification 
further. In the AdS theory without an IR brane, the BR solution is 
replaced by the requirement of regularity at the AdS horizon. In that 
case, the value of $\Phi$ at an arbitrary point with coordinates 
$(x^{\mu}, \theta) $ in the bulk corresponds to $\hat{\lambda}$ 
evaluated at the renormalization scale $\mu = k \; {\rm exp} ( - k r_c 
\theta) $. This then implies that the first order OR differential 
equation for $\Phi$ corresponds to the renormalization group equation 
(RGE) for $\hat{\lambda}$ at energies below the UV cutoff.

In our case, the AdS space has an IR brane and we cannot discard the BR 
solution, which captures the physics associated with the conformal 
symmetry breaking phase transition. However, given that the boundary 
layer has a thickness of just $\sim1/4k r_c\sim\epsilon$ in $\theta$ 
coordinates, as long as we are at least this distance from the IR brane, 
the correspondence between the OR equation for $\Phi$ and the RG 
equation for $\hat{\lambda}$ holds. From Eq.~\eqref{eq:full-OR-equation}, 
the OR equation satisfied by $\Phi$ is
 \begin{eqnarray}
\frac{d\:\log\Phi_{OR}}{d\:(kr_c\theta)}= -\frac{m^2}{4k^2} 
-\frac{\eta}{8\sqrt{k}}\:\frac{\Phi_{OR}}{k^{3/2}}
-\frac{\zeta\:k}{24}\:\frac{\Phi_{OR}^2}{k^3}+... \;\;\;.
\label{RG1}
 \end{eqnarray}
 This result agrees with the equation obtained using conventional 
Hamilton-Jacobi methods, in the limit that 
$m^2/4k^2\ll1$~\cite{de-Boer:1999xf}. 

The corresponding RG equation 
satisfied by the coefficient of the deformation $\hat{\lambda}$ is given in Eqs.~\eqref{eq:g-def} and~\eqref{poly}. From the AdS/CFT dictionary, we have that the 
coordinate in the fifth dimension $\theta$ corresponds to the 
renormalization scale $\mu$ in the $4D$ theory, $k \; {\rm exp} ( - kr_c 
\theta ) = \mu$. While it is tempting to use this to relate the various 
terms in Eq.~\eqref{RG1} and Eq.~\eqref{poly}, we are prevented from 
doing so by the fact that RGEs are in general scheme dependent. However, 
because this restriction does not apply to the lowest order term in an 
RGE, we are able to reproduce the familiar AdS/CFT relation between the 
dimensions of operators in the $4D$ theory and the masses of the 
corresponding scalar fields in 5D, $\Delta = 4+m^2/4k^2=4+\epsilon$ for 
$\epsilon \ll 1$. In addition, in the case when the bulk mass term is 
small so that the potential for the GW scalar is dominated by the cubic 
self-interaction term, we can also equate
 \begin{equation}
c_1 \hat{\lambda} =  - \frac{\eta}{8\sqrt{k}}\:\frac{\Phi_{OR}}{k^{3/2}} \;.
 \end{equation}
%
Comparing the RGE for $\hat{\lambda}$ and the OR differential equation 
for $\Phi$, it is clear that the limit when $\hat{\lambda}$ is small and 
$g(\hat{\lambda})$ is dominated by the constant term corresponds to the 
AdS potential for $\Phi$ being dominated by the mass term. For larger 
values of $\hat{\lambda}$, the linear and higher order terms in 
$g(\hat{\lambda})$ dominate. This corresponds to the cubic and higher 
order self-interaction terms becoming important in the bulk potential 
for $\Phi$.

The potential for the dilaton is to be compared to that for the radion 
in the two cases that we studied in section~\ref{sec:rad-mass}. Consider 
first the case when the self-interactions for $\Phi$ in the bulk can be 
neglected. In this limit, $g(\hat{\lambda})$ in the RG equation for 
$\hat{\lambda}$ is dominated by the constant term. The potential for the 
dilaton using Eq.~\eqref{eq:potential-dilaton} is then given by
 \begin{eqnarray}
V(\chi)= \frac{\chi^4}{4!}
\left[
\kappa_0
-\kappa_1\:\hat{\lambda}\left(\frac{\chi}{f}\right)^{\Delta-4}
\right]\;.
\label{eq:}
 \end{eqnarray}
 The resulting potential for the radion from Eq.~\eqref{eq:rad-potential-1} 
is given by 
 \begin{eqnarray} 
V(\varphi) &=& \frac{k^4}{F^4}\:\varphi^4\left[\tau+\frac{2\alpha v}{ 
F^{\epsilon}}\:\varphi^{\epsilon}\right] 
=\frac{k^4}{F^4}\:\varphi^4\left[\tau+\frac{2\alpha \left(ve^{-\epsilon 
kr_c\pi}\right)}{ 
\left(Fe^{-kr_c\pi}\right)^{\epsilon}}\:\varphi^{\epsilon}\right] \; , 
\label{eq:} 
 \end{eqnarray} 
 where we have rewritten it in a form that is convenient for 
making the comparison. Bearing in mind that the symmetry breaking scale 
$f$ is of order $ke^{-kr_c\pi}$, the duality allows us to relate
 \begin{eqnarray} 
\hat{\lambda} (f) &\Leftrightarrow& 
\frac{\Phi_{OR}(\theta\sim\pi)}{k^{3/2}}\approx ve^{-\epsilon kr_c\pi} 
\nonumber\\ 
\Delta&=&4+\epsilon \;.
\label{eq:} 
 \end{eqnarray}
 With this identification, the potentials for the radion and the dilaton 
are of the same form. The low energy parameter $\kappa_1$ on the CFT 
side is related to the coefficient $\alpha$ of the potential for $\Phi$ 
on the visible brane. Our initial identification $\kappa_0 
\Leftrightarrow \tau$ continues to hold even though $\tau$ now receives 
additional $\mathcal{O}(1)$ contributions from visible brane dynamics. 
Since the potentials for $\varphi$ and $\chi$ are of the same form, the 
leading order expression for the masses of the fluctuations of $\varphi$ 
and $\chi$ have the same parametric dependence.

Consider next the scenario when the potential for the GW scalar $\Phi$ is 
dominated by the cubic self-interaction term in the bulk. By duality 
this is related to the case when $g(\hat{\lambda})$ in the RG equation 
for $\hat{\lambda}$ is dominated by the linear term $c_1\hat{\lambda}$. 
We can read off the potential for the
dilaton from Eq.~\eqref{eq:potential-dilaton}. In this limit, 
 \begin{eqnarray}
V(\chi)= \frac{\chi^4}{4!}
\left[
\kappa_0
-\kappa_1\:\hat{\lambda}\left(\frac{\chi}{f}\right)^{-c_1\hat{\lambda}}
\right] \; .
\label{eq:}
 \end{eqnarray}
 In the corresponding limit, we obtain the potential for the radion 
from Eq.~\eqref{eq:rad-potential-cubic} as
 \begin{align}
V(\varphi)&=&\frac{k^4}{F^4}\:\varphi^4\left[\tau+\frac{2\alpha v}{1-\xi\log(\varphi/F)}\right]
=\frac{k^4}{F^4}\:\varphi^4\left[\tau+\frac{2\alpha v}{1+\xi kr_c\pi-\xi\log(\varphi/Fe^{-kr_c\pi})}\right]\;,
\label{eq:}
 \end{align}
 where $\xi=\eta v/8\sqrt{k}$. In the 
second equality we have added and subtracted $\xi kr_c\pi$ in the 
denominator. For $\xi/(1+\xi kr_c\pi)\ll1$, we can approximate the 
potential as 
 \begin{eqnarray} 
V(\varphi)\approx\varphi^4\left[\tau+\frac{2\alpha v}{1+\xi 
kr_c\pi}\left(\frac{\varphi}{Fe^{-kr_c\pi}}\right)^{\frac{\xi}{1+\xi 
kr_c\pi}}\right] \; . 
\label{eq:} 
 \end{eqnarray}
 This is a more convenient form to compare against the dilaton potential. 
In the limit 
in which we are working, $f$ is of order $ke^{-kr_c\pi}$. The duality 
allows us to relate
 \begin{eqnarray} 
\hat{\lambda}(f) &\Leftrightarrow& \frac{v}{1+\xi kr_c\pi} \nonumber\\ 
c_1\hat{\lambda}(f) &=& -\frac{\xi}{1+\xi kr_c\pi} \;.
\label{eq:} 
 \end{eqnarray} 
 We once again see that the potentials for the radion and the dilaton 
are of the same form. The low energy parameters in the CFT, $\kappa_0$ and 
$\kappa_1$ are again related to the AdS parameters $\tau$ and $\alpha$ 
respectively. The parametric dependences of the masses of the radion and 
the dilaton then agree in a straightforward manner.

Finally, we can use the correspondence from Eq.~\eqref{eq:lambda_corres} to get a better understanding of the behavior of $\Phi(\theta)$ shown in Fig.~\ref{fig:1}.  Since $\theta$ is dual to $\log\mu$, we see that in the outer region, which corresponds to most of the space, the coupling runs logarithmically as we would expect for a nearly marginal deformation of the CFT. Near the IR brane at $\theta=\pi$, the behavior of $\Phi$ changes due to the formation of a boundary layer.  This is the region where the behavior of $\Phi(\theta)$ can no longer be described by the OR Eq.~\eqref{eq:full-OR-equation} and we need a different description.
In the dual picture, this corresponds to the phase transition associated with the spontaneous breakdown of the CFT in the IR, caused by the deformation growing large.
Thus we see that in the outer region, Fig.~\ref{fig:1} is also a good description of the behavior of $\hat{\lambda}(\mu)$, the coefficient of the deformation in the dual theory.

\section{Conclusion}
\label{sec:conclusion}

In this paper we have constructed the effective field theory for a radion which is stabilized by the GW mechanism, but which remains light compared to the KK scale of the extra dimension. Our analysis differs from the original GW construction in that we do not tune the gravitational potential for the radion to vanish, but allow an interplay between the dynamics of the GW scalar and gravity to stabilize the radion. We require that the bulk mass for the GW scalar $\Phi$ is small compared to the inverse curvature scale in order to generate a large hierarchy between the scales of the hidden and visible branes.

We consider two different cases for the bulk potential of the GW scalar. In the first, the mass for $\Phi$, although small, is still the dominant term in the potential. This is the case most often studied in the literature, and we find that the radion is light. In particular, the smallness of the bulk mass in units of the inverse AdS curvature translates to radion being lighter than the KK scale. However, the assumption that the bulk potential is dominated by the mass term is not expected to be satisfied in the general case of theories with a CFT dual. 
Relaxing this assumption, we consider the second case when the bulk potential for $\Phi$ is dominated by a cubic term. This captures the features of a general interacting potential. We take the coefficient of the cubic coupling to be around its strong coupling value. This case differs from the previous one because the radion is generically not light. On the other hand, its mass is controlled by the visible brane tension which can be tuned to allow the radion to lie below the KK scale, even if all the GW interactions are at their strong coupling values. The tuning is mild, scaling as the mass of the radion rather than as the square of the mass.

We have analyzed our results in light of the AdS/CFT correspondence, exploring the connection to the recent results in the literature for a light dilaton. We focused on CFTs which are deformed by marginal operators that are small in the UV, but grow large in the IR to trigger the breaking of conformal symmetry. In general, the scaling behavior of these operators near the breaking scale is very different from their scaling behavior in the UV. The AdS/CFT dictionary associates this change in scaling behavior with the presence of self-interaction terms for the Goldberger-Wise scalar $\Phi$ in the bulk. It follows from this that the inclusion of bulk self-interactions for $\Phi$, such as the cubic coupling considered here, are necessary to describe the properties of the radion in RS theories with holographic duals. Our results for the radion are in good agreement with those obtained earlier for the dilaton. In particular, we find that the presence of a light radion is not a robust prediction of RS models with a holographic dual, but is instead associated with mild tuning. 

Finally, we have analyzed radion couplings to SM fields living on the visible brane, focusing in particular on corrections due to the GW mechanism. These corrections are proportional to the mass squared of the radion in units of the KK scale, and are small if the radion is light. Because in the classical limit the radion coupling to each SM field is proportional to some power of its mass, the corrections from stabilization are subleading for massive SM fields, but can be important and possibly dominant for the photon and the gluon.

\acknowledgments
We would like to thank Raman Sundrum for many invaluable discussions. ZC and RKM are supported by the NSF under grant PHY-0968854. DS is supported in part by the NSF under grant PHY-0910467 and gratefully acknowledges support from the Maryland Center for Fundamental Physics.

\appendix


\section{$5D$ Gravity without a GW Scalar}
\label{app:a}
We give here the intermediate steps in obtaining the effective $4D$ action that includes the $4D$ graviton and the radion fields. In the absence of a stabilization mechanism, the $5D$ action is
\begin{align}
\mathcal{S}_{GR}^{5D}
&=
\int d^4x\int_{-\pi}^{\pi}d\theta\;\sqrt{G}\Big(-2M_5^3\mathcal{R}[G]-\Lambda_b\Big)
-\sqrt{-G_h}\delta(\theta)T_h-\sqrt{-G_v}\delta(\theta-\pi)T_v\;,
\label{eq:GravLag}
\end{align}
where the geometry is a slice of AdS$_5$ compactified on a circle $\mathcal{S}_1$ with $\mathcal{Z}_2$ orbifolding. The metric is then given as 
\begin{eqnarray}
ds^2&=&e^{-2kr(x)|\theta|}g_{\mu\nu}dx^\mu dx^\nu-r^2(x)d\theta^2\;,
\label{eq:}
\end{eqnarray}
where $g_{\mu\nu}$ is the $4D$ graviton and $r(x)$ is the (non-canonical) radion. The $5D$ Ricci scalar for this metric is 
\begin{eqnarray}
\mathcal{R}[G]
&=&
\frac{2}{r(x)}
\left[
e^{2kr(x)|\theta|}
\Big(
-\frac{r}{2}\mathcal{R}[g]-\partial^2 r+3kr|\theta|\partial^2 r+2k|\theta|\partial r\partial r
-3k^2r|\theta|^2\partial r\partial r
\Big)
\right.\nonumber
\\
&&\qquad\qquad\qquad\Big.
+10k^2r-8k\delta(\theta)+8k\delta(\theta-\pi)
\Big]\;.
\label{eq:}
\end{eqnarray}
Using $\sqrt{G}=r(x)e^{-4kr(x)|\theta|}\sqrt{-g}$, the integrand in the action is
\begin{align}
&\sqrt{-g}
e^{-2kr(x)|\theta|}
\left[
2M_5^3
\Big(
r\mathcal{R}[g]
+2\partial^2r
-4k|\theta|\partial r\partial r
-6kr|\theta|\partial^2 r
+6k^2r|\theta|^2\partial r\partial r
\Big)
\right]
\nonumber \\
&\:+
\sqrt{-g}e^{-4kr(x)|\theta|}
\left[-r\Lambda_b+2M_5^3\Big(-20k^2r+16k\delta(\theta)-16k\delta(\theta-\pi)\Big)
-T_h\delta(\theta)-T_v\delta(\theta-\pi)
\right]\;.
\label{eq:}
\end{align}
Using integration by parts,
the integrand further reduces to
\begin{align}
&\sqrt{-g}
e^{-2kr(x)|\theta|}
\left[
2M_5^3
\Big(
r\mathcal{R}[g]
+6k|\theta|\partial r\partial r
-6k^2|\theta|^2r\partial r\partial r
\Big)
\right]
\nonumber \\
&\:+
\sqrt{-g}e^{-4kr(x)|\theta|}
\left[-r\Lambda_b+2M_5^3\Big(-20k^2r+16k\delta(\theta)-16k\delta(\theta-\pi)\Big)
-T_h\delta(\theta)-T_v\delta(\theta-\pi)
\right]\;.
\label{eq:}
\end{align}
Integrating over $\theta$ leads to a cancellation between the $|\theta|$ and the $|\theta|^2$ terms as
\begin{eqnarray}
\int_{-\pi}^{\pi}d\theta e^{-2kr|\theta|}\big(6k|\theta|-6k^2r|\theta|^2\big)
=6k\pi^2e^{-2kr\pi}
\label{eq:}
\end{eqnarray}  
and leads to the effective action for the $4D$ graviton $g_{\mu\nu}(x)$ and the modulus field $r(x)$ as
\begin{eqnarray}
\mathcal{S}_{GR}^{4D}
&=&
\frac{2M_5^3}{k}\int d^4x\sqrt{-g}\Big(1-e^{-2k\pi r(x)}\Big)\mathcal{R}[g]
+
\frac{12M_5^3}{k}\int d^4x\sqrt{-g}
\partial_\mu\Big(e^{-k\pi r(x)}\Big)\partial^\mu\Big(e^{-k\pi r(x)}\Big)
\nonumber\\
&&\qquad
-\int d^4x\sqrt{-g}\;V(r)\;.
\label{eq:}
\end{eqnarray}
The potential $V(r)$ for the modulus field has contributions both from bulk and brane, and is given by
\begin{align}
V(r)
&=
-\int_{-\pi}^{\pi}d\theta e^{-4kr(x)|\theta|}\Big(-r\Lambda_b-40M_5^3k^2r\Big)
+\Big(T_h-32M_5^3k\Big)+\Big(T_v+32M_5^3k\Big)e^{-4kr(x)\pi}
\nonumber\\
&=
e^{-4kr(x)\pi}\bigg(T_v+32M_5^3k-2\frac{r\Lambda_b+40M_5^3k^2r}{4kr}\bigg)+\bigg(T_h-32M_5^3k+2\frac{r\Lambda_b+40M_5^3k^2r}{4kr}\bigg) \;.
\label{eq:}
\end{align}
In terms of the canonically normalized $4D$ radion field $\varphi\equiv Fe^{-k\pi r(x)}$, where $F=\sqrt{24M_5^3/k}$, the $4D$ action looks like
\begin{align}
\mathcal{S}_{GR}^{4D}
&=
2M_5^3/k\int d^4x\sqrt{-g}\Big(1-(\varphi/F)^2\Big)\mathcal{R}
+
\frac12\int d^4x\sqrt{-g}\;
\partial_\mu\varphi\partial^\mu\varphi
\nonumber\\
&\qquad
-\int d^4x\;\sqrt{-g}\left[(\varphi/F)^4\Big(T_v-\frac{\Lambda_b}{2k}+12M_5^3k\Big)+\Big(T_h+\frac{\Lambda_b}{2k}-12M_5^3k\Big)\right]\;.
\label{eq:}
\end{align}
We ignore the interactions between the Ricci scalar and radion because they are small and irrelevant for phenomenology. Using $k^2=-\Lambda_b/24M_5^3$, the radion potential is given as
\begin{eqnarray}
V(\varphi)= \left[(\varphi/F)^4\Big(T_v-\frac{\Lambda_b}{k}\Big)+\Big(T_h+\frac{\Lambda_b}{k}\Big)\right]
\equiv \frac{k^4}{F^4}\left(\tau\:\varphi^4+\varrho\right)\;,
\label{eq:phi-potapp}
\end{eqnarray}
where $\varrho=\Big(T_h+\frac{\Lambda_b}{k}\Big)/k^4$ is the $4D$ cosmological constant which we tune to be small, and $\tau=\Big(T_v-\frac{\Lambda_b}{k}\Big)/k^4$. When the GW scalar is included, $\tau$ will receive additional corrections.

\section{Radion Mass}
\label{app:b}
We give here the intermediate steps involved in obtaining the potential for the radion, the minimization condition and the mass of the radion. We consider two cases for the bulk potential $V_b(\Phi)$: (i) the bulk potential dominated by a mass term and (ii) the bulk potential dominated by a cubic term.  We will give the steps for obtaining the radion potential in both the cases. In the case where the bulk potential only has a mass term, the equation for $\Phi$ admits exact solution. We compare the approximate OR and BR solutions for $\Phi$ obtained earlier to the exact solution of $\Phi$ in this case.
\subsection*{GW Scalar without Bulk Interactions}
Consider first the case of 
\begin{eqnarray}
V_b(\Phi)=\frac12m^2\Phi^2 \;.
\label{eq:}
\end{eqnarray}
The equation satisfied by $\Phi$ in the bulk becomes
\begin{eqnarray}
\partial_\theta^2\Phi-4kr_c\partial_\theta\Phi-m^2r_c^2\Phi=0\;.
\label{eq:}
\end{eqnarray}
The equation being a homogeneous second order differential equation, admits exact solutions given by
\begin{align}
\Phi(\theta,r)&= A e^{\nu_1 kr_c|\theta|}+Be^{\nu_2 kr_c|\theta|}\;,
\label{eq:2.3}
\end{align}
where $\nu_{1,2}=2\pm\sqrt{4+m^2/k^2}$ and $A,B$ are constants, determined by the brane potentials as
\begin{align}
A\nu_1+B\nu_2&=\frac{1}{2k}V_h'(\Phi)
\nonumber\\
\nu_1Ae^{\nu_1kr_c\pi}+\nu_2Be^{\nu_2kr_c\pi}&= -\frac{1}{2k}V_v'(\Phi) 
\label{eq:2.4}
\end{align}
Without specifying the form of $V_h$, we require that the value of $\Phi(\theta=0)=k^{3/2}v$. For the visible brane, we assume a simple form $V_v(\Phi)=2k^{5/2}\alpha\:\Phi$. This gives
\begin{eqnarray}
A+B=k^{3/2}v
\nonumber\\
\nu_1Ae^{\nu_1kr_c\pi}+\nu_2Be^{\nu_2kr_c\pi}&= -k^{3/2}\alpha\;.
\label{eq:}
\end{eqnarray}
We take $\epsilon=m^2/4k^2\ll1$ and $e^{-kr_c\pi}\ll1$ but $e^{\pm\epsilon kr_c\pi}$ unhierarchical. 
The two conditions on $A,B$ fix them to be
\begin{eqnarray}
A&=&k^{3/2}\frac{v\:\nu_2e^{\nu_2kr_c\pi}+\alpha}{\nu_2e^{\nu_2kr_c\pi}-\nu_1e^{\nu_1kr_c\pi}}
\approx -\frac{k^{3/2}\alpha}{4}e^{-(4+\epsilon)kr_c\pi}
\nonumber\\
B&=&k^{3/2}\frac{v\:\nu_1e^{\nu_1kr_c\pi}+\alpha}{\nu_1e^{\nu_1kr_c\pi}-\nu_2e^{\nu_2kr_c\pi}}
\approx k^{3/2}v\;.
\label{eq:}
\end{eqnarray}
Therefore, the leading order solution of $\Phi$ is given as
\begin{eqnarray}
\Phi(\theta)=-\frac{k^{3/2}\alpha}{4}e^{(4+\epsilon)kr_c(\theta-\pi)}
+k^{3/2}v e^{-\epsilon kr_c\theta}\;.
\label{eq:}
\end{eqnarray}
Using the arguments from section~\ref{sec:rad-mass}, an approximate solution for $\Phi$ can be obtained by a boundary layer analysis:
\begin{eqnarray}
\Phi_{\rm approx}(\theta)= -\frac{k^{3/2}\alpha}{4}e^{4kr_c(\theta-\pi)}
+k^{3/2}v e^{-\epsilon kr_c\theta}\;.
\label{eq:}
\end{eqnarray}
Figure~\ref{fig:1} in section~\ref{sec:rad-mass} shows that the approximate solution agrees very well with the exact solution.

\subsubsection*{Radion Potential}

Substituting either solution of $\Phi$ in the action, integrating over the extra dimension and letting $r_c\rightarrow r(x)$ generates a potential for the radion. The potential gets contribution both from the bulk and the brane terms in the GW action, and is given in general by
\begin{eqnarray}
V_{GW}(r)
&=&
\int_{0}^{\pi} d\theta\:\frac{1}{r}e^{-4kr\theta}
\Big(
\partial_\theta\Phi\;\partial_\theta\Phi
+r^2m^2\Phi^2
\Big)
+ e^{-4kr\pi} 2\alpha k^{5/2}\Phi(\pi)\;.
\label{eq:radion-pot}
\end{eqnarray}
Working in the limit where $|\epsilon|\ll 1$, the bulk contribution to the potential goes as
\begin{align}
V_{GW}^{\rm bulk}(r)
&=k\left[(4+\epsilon)A^2\bigg(e^{(2\nu_1-4)kr\pi}-1\bigg)
-\epsilon B^2\bigg(e^{(2\nu_2-4)kr\pi}-1\bigg)
\right]
\nonumber\\
&=
k^4\:\frac{e^{-4kr\pi}}{4+\epsilon}
\Big(\alpha-\epsilon\:ve^{-\epsilon kr_c\pi}\Big)^2
-\epsilon\:k^4\:v^2 e^{-(4+2\epsilon) kr_c\pi}
+\epsilon\:k^4\:v^2\;.
\label{eq:}
\end{align}
The brane contribution to the potential is
\begin{align}
V_{GW}^{\rm brane}(r)&=
2k^{5/2}\:\alpha\:e^{-4kr\pi}\Phi(\pi)
\nonumber\\
&=
k^4\:e^{-4kr\pi}
2\alpha
\left[
-\frac{\alpha-\epsilon\:ve^{-\epsilon kr_c\pi}}{4+\epsilon}
+ve^{-\epsilon kr_c\pi}
\right]\;.
\label{eq:brane-pot}
\end{align}
Keeping to linear order in $v$ and $\epsilon$ we find that the bulk contribution to the potential is subleading. We can now write the leading radion potential, including the gravitational contributions from Eq.~\eqref{eq:phi-potapp}, in terms of the canonical field $\varphi$:
\begin{eqnarray}
V(\varphi)
&=& k^4\:\Big(\frac{\varphi}{F}\Big)^4
\left[
\tau
+2\alpha\:v\:\Big(\frac{\varphi}{F}\Big)^\epsilon
\right]\;.
\label{eq:}
\end{eqnarray}
Here, $\tau$ receives contributions from both the gravity sector and the visible brane dynamics, and, to leading order in $\epsilon$, it is given by
\begin{eqnarray}
\tau=\frac{T_v-\Lambda_b/k-k^4 \alpha^2/4}{k^4}\;.
\label{eq:}
\end{eqnarray}
The minimization condition gives $\left<\varphi\right>=f$ as
\begin{eqnarray}
\tau+2\,\alpha\,v\,\left(\frac{f}{F}\right)^{\epsilon}
=0\;.
\end{eqnarray}
At the minimum, the mass squared is given as
\begin{eqnarray}
m_\varphi^2 &=&-\frac{\epsilon\,\tau}{6}\frac{k^3}{M_5^3}\left(ke^{-kr_c\pi}\right)^2.
\label{eq:}
\end{eqnarray} 
Including higher order terms does not change the parametric dependences.

\subsection*{GW Scalar with Bulk Cubic Self Interaction}
We next consider the case of 
\begin{eqnarray}
V_b(\Phi)=\frac{1}{3!}\eta\:\Phi^3\;.
\label{eq:}
\end{eqnarray}
In this case, the classical equation satisfied by $\Phi$ is given as
\begin{eqnarray}
\partial_\theta^2\Phi-4kr_c\partial_\theta\Phi-\frac{\eta}{2}\:r_c^2\Phi^2=0\;.
\label{eq:}
\end{eqnarray}
In the absence of an exact solution, we use the boundary layer theory described in section~\ref{sec:radion_mass} to get the approximate solution for $\Phi$ as
\begin{eqnarray}
\Phi_{\rm approx}(\theta)=
-\frac{k^{3/2}\alpha}{4}e^{4kr_c(\theta-\pi)}
+\frac{k^{3/2}v}{1+\xi kr\theta} \;,
\label{eq:cubic-phi}
\end{eqnarray}
where $\xi=\eta\:v/8\sqrt{k}$.

The radion potential coming from the GW action with a cubic interaction in the bulk is given by
\begin{eqnarray}
V_{GW}(r)
&=&
\int_{0}^{\pi} d\theta\:\frac{1}{r}e^{-4kr\theta}
\Big(
\partial_\theta\Phi\;\partial_\theta\Phi
+\frac{r^2\,\eta}{3}\Phi^3
\Big)
+ e^{-4kr\pi} 2\alpha k^{5/2}\Phi(\pi)\;.
\label{eq:radion-pot-int}
\end{eqnarray}
We now plug the solution from Eq.~\eqref{eq:cubic-phi} into Eq.~\eqref{eq:radion-pot-int} to get an explicit form of the radion potential. Including the gravitational contribution and working to leading order in $v$ we get
\begin{eqnarray}
V(\varphi)=k^4\left(\frac\varphi F\right)^4
\left[
\tau+\frac{1}{1-\xi\log(\varphi/F)}\left( 2\alpha v+\frac{\alpha^2\xi}{8} \right)
\right] \; .
\label{eq:}
\end{eqnarray}
We define $w= 2\alpha v+\alpha^2\xi/8$. The first term in $w$ comes from the brane potential while the second term comes from the bulk, and we note that both are small because they are proportional to $v$. 

The potential is minimized at $\left<\varphi\right>=f$ as
\begin{eqnarray}
\tau+\frac{w}{1-\xi\log(f/F)}=0 \; .
\label{eq:}
\end{eqnarray}
The mass squared at the minimum is given by
\begin{eqnarray}
m_\varphi^2 &=&\frac{\tau^2\xi}{6w  }\, \frac{k^3}{M_5^3}
\left(ke^{-kr_c\pi}\right)^2 .
\label{eq:}
\end{eqnarray}
At the NDA values of the parameters (see the next section), we find that $m_\varphi^2/m_{KK}^2\sim\tau^2$.

\section{NDA Estimation of Parameters}
\label{app:NDA}

In this appendix we give a brief review of estimates using naive 
dimensional analysis (NDA)~\cite{Manohar:1983md,Georgi:1986kr} and its 
application in five dimensions~\cite{Chacko:1999hg} as in this work. The 
general idea of NDA is to estimate the size of Lagrangian parameters by 
assuming that quantum corrections are the same size at every loop order.

The loop factor $\ell_D$ that comes from integrating over 
$D$-dimensional phase space is given by
 \begin{equation}
\ell_D = 2^D \pi^{D/2} \Gamma(D/2)\;,
 \end{equation} 
 giving the familiar $\ell_4 = 16 \pi^2$. The relevant number for five 
dimensions is $\ell_5 = 24 \pi^3$. If there are $N$ states in the theory 
which go around in loops, then each loop contribution gets multiplied by 
$N$. For our case $N$ is small, of order a few. We can then write the 
$D$ dimensional Lagrangian as follows
 \begin{equation}
{\cal L}_D  \sim \frac{N \Lambda^D}{\ell_D} \hat{{\cal L}}(\hat{\Phi}, \partial/\Lambda)\;,
\label{eq:NDA_Lag}
 \end{equation} 
 where $\hat{\Phi}$ is a field whose kinetic term is not canonically 
normalized, $\Lambda$ is the cutoff of the theory, and all parameters in 
$\hat{\cal L}$ are dimensionless and $O(1)$. 

\subsection*{Gravity Lagrangian}

We can begin with the gravity Lagrangian of Eq.~\eqref{eq:GravLag} and 
use the fact that the kinetic term has two derivatives to estimate the 
cutoff:
 \begin{equation}
\Lambda \sim \left( \frac{\ell_5}{N}\right)^{1/3} M_5 = \left(\frac{3}{N}\right)^{1/3}\, 2\pi\,  M_5\;,
 \end{equation}
 where $M_5$ is the five dimensional Planck mass. From this we see that 
there is a regime of the effective field theory before 5$D$ gravity 
becomes strongly coupled if $N$ is not too large. Next we estimate the 
natural value of the cosmological constant $\Lambda_b$
 \begin{equation}
\Lambda_{ b} \sim \frac{N \Lambda^5}{\ell_5} \sim \left( \frac{\ell_{5}}{N} \right)^{2/3} M_5^5\;,
 \end{equation}
 from which we can estimate the inverse of the AdS curvature $k$,
 \begin{equation}
k = \sqrt{\frac{-\Lambda_{ b}}{24 M_5^3}} \sim \frac{1}{2\sqrt{6}} \left(\frac{\ell_5}{N} \right)^{1/3}  M_5 \sim \frac{2}{N^{1/3}} M_5\sim \frac{\Lambda}{\sqrt{24}}\;.
 \end{equation}
 From this we see that the AdS curvature scale is only separated by a 
factor of a few from the cutoff $\Lambda$ and the higher dimensional 
Planck scale $M_5$. This implies that if the bulk cosmological constant 
is of order its natural value, only a handful of KK states are present 
below the cutoff.

The AdS/CFT correspondence assumes the hierarchy $M_5 \gg M_S \gg k$. 
Here $M_S$ represents the string scale, the energy scale at which string 
excitations enter the picture. From the CFT perspective, this 
corresponds to requiring that $N_c \gg 1$ and $g^2 N_c \gg 1$, where 
$N_c$ is the number of colors and $g$ the coupling constant in the dual 
gauge theory. The fact that $M_5$ and $k$ differ only by a factor of a 
few implies that we are not actually in the regime where $N_c$ and $g^2 
N_c$ are large. This implies that the results we obtain using our NDA 
estimates can only be taken as a very rough guide, since the 
correspondence is being pushed to the edge of its domain of validity.

We can also use NDA to estimate the natural values of the four 
dimensional cosmological constants, which in this case are the brane 
tensions:
 \begin{equation}
T_h \sim T_v \sim \frac{N\Lambda^4}{\ell_4} \sim \frac{\ell_5^{4/3}}{\ell_4 \,N^{1/3}} M_5^4\;.
 \end{equation}
 These parameters are restricted to a four dimensional brane, so it is the 
four dimensional loop factor which goes into the estimate.

\subsection*{GW Field}

The next step is to estimate the values of the parameters in the 
potential of the GW field, $\Phi$. We begin with the bulk parameters 
defined in Eq.~\eqref{eq:bulk_pot}. In order to get from the Lagrangian 
in Eq.~\eqref{eq:NDA_Lag}, to one with field that have canonical kinetic 
terms, we have to rescale $\Phi$ by $\sqrt{\ell_5/ N\Lambda^{3}}$.  
Therefore, for a mass term $m^2 \Phi^2/2$, the natural value of the 
mass is given by
 \begin{equation}
m^2  \sim \Lambda^2 \sim 24 k^2\;.
 \end{equation}
In order to generate a large hierarchy, the mass parameter is taken to 
lie significantly below its NDA value.  
The natural value of the cubic interaction is also easily obtained as
 \begin{equation}
\eta \sim \sqrt{\frac{\ell_5 \Lambda}{N}} \sim 24^{1/4} \sqrt{\frac{\ell_5 k}{N}} \sim 60 \sqrt{\frac{k}{N}}\;.
 \end{equation}

We also want to compute the natural values of the parameters $\alpha$ 
and $v$.  The visible brane potential for the GW field is given by
 \begin{equation}
V_v = \delta(\theta - \pi) 2 k^{5/2} \alpha \Phi\;.
 \end{equation}
Using the NDA prescription, we estimate the size of this term to be
 \begin{equation}
2\alpha k^{5/2} \sim \frac{\sqrt{N \ell_5}}{\ell_4} \Lambda^{5/2}\sim \frac{\ell_5^{4/3}}{\ell_4 \, N^{1/3}}M_5^{5/2}
 \end{equation}
 \begin{equation}
\alpha \sim 2^{11/4} 3^{5/4} \frac{\sqrt{N\ell_5}}{\ell_4} \sim 3 \left( \frac{54}{\pi^2} \right)^{1/4}\sqrt{N} \sim 5\sqrt{N}\;.
 \end{equation}
 As discussed in section~\ref{sec:radion_mass}, we only require that 
$\Phi(\theta = 0) = k^{3/2}v$, but leave the potential unspecified. One potential that can generate this boundary condition is
 \begin{equation}
V_h = \delta(\theta) \lambda \left(\Phi^2 - k^3 v^2 \right)^2,
 \end{equation}
from which we estimate the natural value of $v$ as
 \begin{equation}
v \sim  24^{3/4}\left(\frac{N}{\ell_5}\right)^{1/2} = \frac{24^{1/4}}{\pi^{3/2}} \sqrt{N} \sim 0.4\sqrt{N}\;.
 \end{equation}
 Using other possible potentials such as $\lambda (\Phi-k^{3/2} v)^2$ 
give the same estimate for $v$. In order to generate a hierarchy we take $v$
to lie below its NDA value.

\subsection*{Radion Dynamics}

The radion parameter $\tau$ associated with the quartic is defined below 
Eq.~\eqref{eq:rad-potential-1} and can be estimated as follows:
 \begin{equation}
\tau = \frac{1}{k^4} \left(T_v - \frac{\Lambda_{b}}{k} -\frac{k^4\alpha^2}{4}\right) \sim 
           72 \,N\left(  \frac{8}{\ell_4} - \frac{16\sqrt{6}}{\ell_5} - \frac{\sqrt{6}\,\ell_5}{\ell_4^2} \right) \sim 
           5\, N.
\label{eq:quarticNDA}
\end{equation}
 The three contributions we show are roughly the same size. Since this is 
just an estimate, there are $O(1)$ coefficients on each term, so we 
assume there is no cancellation and that total size of $\tau$ is the size of each individual term.  

We have computed the mass of the radion in the regime where the OR 
Eq.~\eqref{eq:full-OR-equation} is dominated by $m^2$ and by $\eta$ in 
Eqs.~\eqref{eq:rad_mass_m} and~\eqref{eq:rad_mass_cubic} respectively:
 \begin{equation} \frac{m_\varphi^2}{\left(k e^{-k \pi r_c} 
\right)^2} =
 \left\{ 
  \begin{array}{l l}
    -\frac{\epsilon\tau}{6}\frac{k^3}{M_5^3}  & \quad \text{if $m^2$ dominates}\\[10 pt]
     \frac{\tau^2\xi}{6w} \frac{k^3}{M_5^3}
\sim 10 & \quad \text{if $\eta$ dominates.}
  \end{array} \right.
\end{equation}
These formulae require that the mass term be well below its natural value, so we only give an NDA estimate for the second case in which the radion mass does not depend on $\epsilon$. As the mass of the KK gravitons is typically $\sim 3$ times larger than $k 
e^{-k \pi r_c}$, we see that in the cubic case NDA gives us the expected result that the radion mass is roughly equal to the KK scale.

\subsection*{Radion Coupling to SM}

Finally we estimate the coupling of the GW field to other Standard Model fields on the visible brane. These couplings take the form
\begin{equation}
\left(1+\alpha_{\rm int} \Phi/k^{3/2}\right) \cdot {\cal O}_{\rm SM}\;.
\end{equation}
We can estimate the size of $\alpha_{\rm int}$ using NDA and assuming that the operator ${\cal O}_{\rm SM}$ has already been normalized, so we just need to rescale $\Phi$ to get a canonical kinetic term:
\begin{equation}
\alpha_{\rm int}/k^{3/2} \sim \ell_5^{1/2}/N^{1/2}\Lambda^{3/2}
\end{equation}
\begin{equation}
\alpha_{\rm int} \sim \frac{1}{24^{3/4}}\sqrt{\frac{\ell_5}{N}}\sim \frac{1}{\sqrt{N}}\left( \frac{\pi^6}{24} \right)^{1/4} \sim \frac{2.5}{\sqrt{N}}\;.
\end{equation}
We will take these parameters to be their NDA sizes, but since they will always multiply $\Phi/k^{3/2}$, we will end up only working to first order in them. 

\bibliographystyle{JHEP}
\bibliography{thebigbibfile}

\providecommand{\href}[2]{#2}\begingroup\raggedright\begin{thebibliography}{10}

\bibitem{:2012gu}
{\bf CMS Collaboration} Collaboration, S.~Chatrchyan {\em et~al.}, {\it
  {Observation of a new boson at a mass of 125 GeV with the CMS experiment at
  the LHC}},  {\em Phys.Lett.} {\bf B716} (2012) 30--61,
  [\href{http://arxiv.org/abs/1207.7235}{{\tt arXiv:1207.7235}}].

\bibitem{:2012gk}
{\bf ATLAS Collaboration} Collaboration, G.~Aad {\em et~al.}, {\it {Observation
  of a new particle in the search for the Standard Model Higgs boson with the
  ATLAS detector at the LHC}},  {\em Phys.Lett.} {\bf B716} (2012) 1--29,
  [\href{http://arxiv.org/abs/1207.7214}{{\tt arXiv:1207.7214}}].

\bibitem{Susskind:1978ms}
L.~Susskind, {\it {Dynamics of Spontaneous Symmetry Breaking in the
  Weinberg-Salam Theory}},  {\em Phys.Rev.} {\bf D20} (1979) 2619--2625.

\bibitem{'tHooft:1980xb}
e.~'t~Hooft, Gerard, e.~Itzykson, C., e.~Jaffe, A., e.~Lehmann, H., e.~Mitter,
  P.K., {\em et~al.}, {\it {Recent Developments in Gauge Theories. Proceedings,
  Nato Advanced Study Institute, Cargese, France, August 26 - September 8,
  1979}},  {\em NATO Adv.Study Inst.Ser.B Phys.} {\bf 59} (1980) pp.1--438.

\bibitem{Randall:1999ee}
L.~Randall and R.~Sundrum, {\it {A Large mass hierarchy from a small extra
  dimension}},  {\em Phys.Rev.Lett.} {\bf 83} (1999) 3370--3373,
  [\href{http://arxiv.org/abs/hep-ph/9905221}{{\tt hep-ph/9905221}}].

\bibitem{Agashe:2003zs}
K.~Agashe, A.~Delgado, M.~J. May, and R.~Sundrum, {\it {RS1, custodial isospin
  and precision tests}},  {\em JHEP} {\bf 0308} (2003) 050,
  [\href{http://arxiv.org/abs/hep-ph/0308036}{{\tt hep-ph/0308036}}].

\bibitem{Agashe:2006at}
K.~Agashe, R.~Contino, L.~Da~Rold, and A.~Pomarol, {\it {A Custodial symmetry
  for Zb anti-b}},  {\em Phys.Lett.} {\bf B641} (2006) 62--66,
  [\href{http://arxiv.org/abs/hep-ph/0605341}{{\tt hep-ph/0605341}}].

\bibitem{Csaki:2003zu}
C.~Csaki, C.~Grojean, L.~Pilo, and J.~Terning, {\it {Towards a realistic model
  of Higgsless electroweak symmetry breaking}},  {\em Phys.Rev.Lett.} {\bf 92}
  (2004) 101802, [\href{http://arxiv.org/abs/hep-ph/0308038}{{\tt
  hep-ph/0308038}}].

\bibitem{Contino:2003ve}
R.~Contino, Y.~Nomura, and A.~Pomarol, {\it {Higgs as a holographic
  pseudoGoldstone boson}},  {\em Nucl.Phys.} {\bf B671} (2003) 148--174,
  [\href{http://arxiv.org/abs/hep-ph/0306259}{{\tt hep-ph/0306259}}].

\bibitem{Agashe:2004rs}
K.~Agashe, R.~Contino, and A.~Pomarol, {\it {The Minimal composite Higgs
  model}},  {\em Nucl.Phys.} {\bf B719} (2005) 165--187,
  [\href{http://arxiv.org/abs/hep-ph/0412089}{{\tt hep-ph/0412089}}].

\bibitem{Grossman:1999ra}
Y.~Grossman and M.~Neubert, {\it {Neutrino masses and mixings in
  nonfactorizable geometry}},  {\em Phys.Lett.} {\bf B474} (2000) 361--371,
  [\href{http://arxiv.org/abs/hep-ph/9912408}{{\tt hep-ph/9912408}}].

\bibitem{Gherghetta:2000qt}
T.~Gherghetta and A.~Pomarol, {\it {Bulk fields and supersymmetry in a slice of
  AdS}},  {\em Nucl.Phys.} {\bf B586} (2000) 141--162,
  [\href{http://arxiv.org/abs/hep-ph/0003129}{{\tt hep-ph/0003129}}].

\bibitem{Agashe:2004cp}
K.~Agashe, G.~Perez, and A.~Soni, {\it {Flavor structure of warped extra
  dimension models}},  {\em Phys.Rev.} {\bf D71} (2005) 016002,
  [\href{http://arxiv.org/abs/hep-ph/0408134}{{\tt hep-ph/0408134}}].

\bibitem{Agashe:2004ci}
K.~Agashe and G.~Servant, {\it {Warped unification, proton stability and dark
  matter}},  {\em Phys.Rev.Lett.} {\bf 93} (2004) 231805,
  [\href{http://arxiv.org/abs/hep-ph/0403143}{{\tt hep-ph/0403143}}].

\bibitem{Agashe:2004bm}
K.~Agashe and G.~Servant, {\it {Baryon number in warped GUTs: Model building
  and (dark matter related) phenomenology}},  {\em JCAP} {\bf 0502} (2005) 002,
  [\href{http://arxiv.org/abs/hep-ph/0411254}{{\tt hep-ph/0411254}}].

\bibitem{Medina:2011qc}
A.~D. Medina and E.~Ponton, {\it {Warped Radion Dark Matter}},  {\em JHEP} {\bf
  1109} (2011) 016, [\href{http://arxiv.org/abs/1104.4124}{{\tt
  arXiv:1104.4124}}].

\bibitem{Agashe:2006hk}
K.~Agashe, A.~Belyaev, T.~Krupovnickas, G.~Perez, and J.~Virzi, {\it {LHC
  Signals from Warped Extra Dimensions}},  {\em Phys.Rev.} {\bf D77} (2008)
  015003, [\href{http://arxiv.org/abs/hep-ph/0612015}{{\tt hep-ph/0612015}}].

\bibitem{Contino:2006nn}
R.~Contino, T.~Kramer, M.~Son, and R.~Sundrum, {\it {Warped/composite
  phenomenology simplified}},  {\em JHEP} {\bf 0705} (2007) 074,
  [\href{http://arxiv.org/abs/hep-ph/0612180}{{\tt hep-ph/0612180}}].

\bibitem{Lillie:2007yh}
B.~Lillie, L.~Randall, and L.-T. Wang, {\it {The Bulk RS KK-gluon at the LHC}},
   {\em JHEP} {\bf 0709} (2007) 074,
  [\href{http://arxiv.org/abs/hep-ph/0701166}{{\tt hep-ph/0701166}}].

\bibitem{Goldberger:1999uk}
W.~D. Goldberger and M.~B. Wise, {\it {Modulus stabilization with bulk
  fields}},  {\em Phys.Rev.Lett.} {\bf 83} (1999) 4922--4925,
  [\href{http://arxiv.org/abs/hep-ph/9907447}{{\tt hep-ph/9907447}}].

\bibitem{Csaki:1999mp}
C.~Csaki, M.~Graesser, L.~Randall, and J.~Terning, {\it {Cosmology of brane
  models with radion stabilization}},  {\em Phys.Rev.} {\bf D62} (2000) 045015,
  [\href{http://arxiv.org/abs/hep-ph/9911406}{{\tt hep-ph/9911406}}].

\bibitem{Goldberger:1999un}
W.~D. Goldberger and M.~B. Wise, {\it {Phenomenology of a stabilized modulus}},
   {\em Phys.Lett.} {\bf B475} (2000) 275--279,
  [\href{http://arxiv.org/abs/hep-ph/9911457}{{\tt hep-ph/9911457}}].

\bibitem{Giudice:2000av}
G.~F. Giudice, R.~Rattazzi, and J.~D. Wells, {\it {Graviscalars from higher
  dimensional metrics and curvature Higgs mixing}},  {\em Nucl.Phys.} {\bf
  B595} (2001) 250--276, [\href{http://arxiv.org/abs/hep-ph/0002178}{{\tt
  hep-ph/0002178}}].

\bibitem{Csaki:2000zn}
C.~Csaki, M.~L. Graesser, and G.~D. Kribs, {\it {Radion dynamics and
  electroweak physics}},  {\em Phys.Rev.} {\bf D63} (2001) 065002,
  [\href{http://arxiv.org/abs/hep-th/0008151}{{\tt hep-th/0008151}}].

\bibitem{Csaki:2007ns}
C.~Csaki, J.~Hubisz, and S.~J. Lee, {\it {Radion phenomenology in realistic
  warped space models}},  {\em Phys.Rev.} {\bf D76} (2007) 125015,
  [\href{http://arxiv.org/abs/0705.3844}{{\tt arXiv:0705.3844}}].

\bibitem{Rizzo:2002pq}
T.~G. Rizzo, {\it {Radion couplings to bulk fields in the Randall-Sundrum
  model}},  {\em JHEP} {\bf 0206} (2002) 056,
  [\href{http://arxiv.org/abs/hep-ph/0205242}{{\tt hep-ph/0205242}}].

\bibitem{Maldacena:1997re}
J.~M. Maldacena, {\it {The Large N limit of superconformal field theories and
  supergravity}},  {\em Adv.Theor.Math.Phys.} {\bf 2} (1998) 231--252,
  [\href{http://arxiv.org/abs/hep-th/9711200}{{\tt hep-th/9711200}}].

\bibitem{Witten:1998qj}
E.~Witten, {\it {Anti-de Sitter space and holography}},  {\em
  Adv.Theor.Math.Phys.} {\bf 2} (1998) 253--291,
  [\href{http://arxiv.org/abs/hep-th/9802150}{{\tt hep-th/9802150}}].

\bibitem{Gubser:1998bc}
S.~Gubser, I.~R. Klebanov, and A.~M. Polyakov, {\it {Gauge theory correlators
  from noncritical string theory}},  {\em Phys.Lett.} {\bf B428} (1998)
  105--114, [\href{http://arxiv.org/abs/hep-th/9802109}{{\tt hep-th/9802109}}].

\bibitem{Klebanov:1999tb}
I.~R. Klebanov and E.~Witten, {\it {AdS / CFT correspondence and symmetry
  breaking}},  {\em Nucl.Phys.} {\bf B556} (1999) 89--114,
  [\href{http://arxiv.org/abs/hep-th/9905104}{{\tt hep-th/9905104}}].

\bibitem{ArkaniHamed:2000ds}
N.~Arkani-Hamed, M.~Porrati, and L.~Randall, {\it {Holography and
  phenomenology}},  {\em JHEP} {\bf 0108} (2001) 017,
  [\href{http://arxiv.org/abs/hep-th/0012148}{{\tt hep-th/0012148}}].

\bibitem{Rattazzi:2000hs}
R.~Rattazzi and A.~Zaffaroni, {\it {Comments on the holographic picture of the
  Randall-Sundrum model}},  {\em JHEP} {\bf 0104} (2001) 021,
  [\href{http://arxiv.org/abs/hep-th/0012248}{{\tt hep-th/0012248}}].

\bibitem{Salam:1970qk}
A.~Salam and J.~Strathdee, {\it {Nonlinear realizations. 2. Conformal
  symmetry}},  {\em Phys.Rev.} {\bf 184} (1969) 1760--1768.

\bibitem{Isham:1970gz}
C.~Isham, A.~Salam, and J.~Strathdee, {\it {Spontaneous breakdown of conformal
  symmetry}},  {\em Phys.Lett.} {\bf B31} (1970) 300--302.

\bibitem{Isham:1971dv}
C.~Isham, A.~Salam, and J.~Strathdee, {\it {Nonlinear realizations of
  space-time symmetries. Scalar and tensor gravity}},  {\em Annals Phys.} {\bf
  62} (1971) 98--119.

\bibitem{Zumino:1970ab}
B.~Zumino, {\it Lectures on elementary particles and quantum field theory},  in
  {\em 1970 Brandeis Summer School} (S.~Deser, ed.).
\newblock MIT Press, 1970.

\bibitem{Ellis:1970yd}
J.~R. Ellis, {\it {Aspects of conformal symmetry and chirality}},  {\em
  Nucl.Phys.} {\bf B22} (1970) 478--492.

\bibitem{Ellis:1971sa}
J.~R. Ellis, {\it {Phenomenological actions for spontaneously-broken conformal
  symmetry}},  {\em Nucl.Phys.} {\bf B26} (1971) 536--546.

\bibitem{Goldberger:2007zk}
W.~D. Goldberger, B.~Grinstein, and W.~Skiba, {\it {Distinguishing the Higgs
  boson from the dilaton at the Large Hadron Collider}},  {\em Phys.Rev.Lett.}
  {\bf 100} (2008) 111802, [\href{http://arxiv.org/abs/0708.1463}{{\tt
  arXiv:0708.1463}}].

\bibitem{Fan:2008jk}
J.~Fan, W.~D. Goldberger, A.~Ross, and W.~Skiba, {\it {Standard Model couplings
  and collider signatures of a light scalar}},  {\em Phys.Rev.} {\bf D79}
  (2009) 035017, [\href{http://arxiv.org/abs/0803.2040}{{\tt
  arXiv:0803.2040}}].

\bibitem{Vecchi:2010gj}
L.~Vecchi, {\it {Phenomenology of a light scalar: the dilaton}},  {\em
  Phys.Rev.} {\bf D82} (2010) 076009,
  [\href{http://arxiv.org/abs/1002.1721}{{\tt arXiv:1002.1721}}].

\bibitem{RattazziPlanck:2010}
R.~Rattazzi, {\it {The Naturally Light Dilaton}}, . Planck 2010.

\bibitem{Chacko:2012sy}
Z.~Chacko and R.~K. Mishra, {\it {Effective Theory of a Light Dilaton}},  {\em
  Phys.Rev.} {\bf D87} (2013) 115006,
  [\href{http://arxiv.org/abs/1209.3022}{{\tt arXiv:1209.3022}}].

\bibitem{Bellazzini:2012vz}
B.~Bellazzini, C.~Csaki, J.~Hubisz, J.~Serra, and J.~Terning, {\it {A Higgslike
  Dilaton}},  {\em Eur.Phys.J.} {\bf C73} (2013) 2333,
  [\href{http://arxiv.org/abs/1209.3299}{{\tt arXiv:1209.3299}}].

\bibitem{Abe:2012eu}
T.~Abe, R.~Kitano, Y.~Konishi, K.-y. Oda, J.~Sato, {\em et~al.}, {\it {Minimal
  Dilaton Model}},  {\em Phys.Rev.} {\bf D86} (2012) 115016,
  [\href{http://arxiv.org/abs/1209.4544}{{\tt arXiv:1209.4544}}].

\bibitem{Coradeschi:2013gda}
F.~Coradeschi, P.~Lodone, D.~Pappadopulo, R.~Rattazzi, and L.~Vitale, {\it {A
  naturally light dilaton}},  \href{http://arxiv.org/abs/1306.4601}{{\tt
  arXiv:1306.4601}}.

\bibitem{Charmousis:1999rg}
C.~Charmousis, R.~Gregory, and V.~Rubakov, {\it {Wave function of the radion in
  a brane world}},  {\em Phys.Rev.} {\bf D62} (2000) 067505,
  [\href{http://arxiv.org/abs/hep-th/9912160}{{\tt hep-th/9912160}}].

\bibitem{Luty:2000ec}
M.~A. Luty and R.~Sundrum, {\it {Hierarchy stabilization in warped
  supersymmetry}},  {\em Phys.Rev.} {\bf D64} (2001) 065012,
  [\href{http://arxiv.org/abs/hep-th/0012158}{{\tt hep-th/0012158}}].

\bibitem{Breitenlohner:1982jf}
P.~Breitenlohner and D.~Z. Freedman, {\it {Stability in Gauged Extended
  Supergravity}},  {\em Annals Phys.} {\bf 144} (1982) 249.

\bibitem{benOrz}
C.~M. Bender and S.~A. Orszag, {\em {Advanced Mathematical methods for
  Scientists and Engineers}}.
\newblock McGraw-Hill Publishing Company, 1978.

\bibitem{Konstandin:2010cd}
T.~Konstandin, G.~Nardini, and M.~Quiros, {\it {Gravitational Backreaction
  Effects on the Holographic Phase Transition}},  {\em Phys.Rev.} {\bf D82}
  (2010) 083513, [\href{http://arxiv.org/abs/1007.1468}{{\tt
  arXiv:1007.1468}}].

\bibitem{Eshel:2011wz}
Y.~Eshel, S.~J. Lee, G.~Perez, and Y.~Soreq, {\it {Shining Flavor and Radion
  Phenomenology in Warped Extra Dimension}},  {\em JHEP} {\bf 1110} (2011) 015,
  [\href{http://arxiv.org/abs/1106.6218}{{\tt arXiv:1106.6218}}].

\bibitem{Georgi:1974yw}
H.~Georgi and A.~Pais, {\it {Calculability and Naturalness in Gauge Theories}},
   {\em Phys.Rev.} {\bf D10} (1974) 539.

\bibitem{Georgi:1975tz}
H.~Georgi and A.~Pais, {\it {Vacuum Symmetry and the PseudoGoldstone
  Phenomenon}},  {\em Phys.Rev.} {\bf D12} (1975) 508.

\bibitem{Kaplan:1983fs}
D.~B. Kaplan and H.~Georgi, {\it {SU(2) x U(1) Breaking by Vacuum
  Misalignment}},  {\em Phys.Lett.} {\bf B136} (1984) 183.

\bibitem{Kaplan:1983sm}
D.~B. Kaplan, H.~Georgi, and S.~Dimopoulos, {\it {Composite Higgs Scalars}},
  {\em Phys.Lett.} {\bf B136} (1984) 187.

\bibitem{Georgi:1984af}
H.~Georgi and D.~B. Kaplan, {\it {Composite Higgs and Custodial SU(2)}},  {\em
  Phys.Lett.} {\bf B145} (1984) 216.

\bibitem{Arzt:1993gz}
C.~Arzt, {\it {Reduced effective Lagrangians}},  {\em Phys.Lett.} {\bf B342}
  (1995) 189--195, [\href{http://arxiv.org/abs/hep-ph/9304230}{{\tt
  hep-ph/9304230}}].

\bibitem{Georgi:1991ch}
H.~Georgi, {\it {On-shell effective field theory}},  {\em Nucl.Phys.} {\bf
  B361} (1991) 339--350.

\bibitem{Porrati:1999ew}
M.~Porrati and A.~Starinets, {\it {RG fixed points in supergravity duals of 4-D
  field theory and asymptotically AdS spaces}},  {\em Phys.Lett.} {\bf B454}
  (1999) 77--83, [\href{http://arxiv.org/abs/hep-th/9903085}{{\tt
  hep-th/9903085}}].

\bibitem{Balasubramanian:1999jd}
V.~Balasubramanian and P.~Kraus, {\it {Space-time and the holographic
  renormalization group}},  {\em Phys.Rev.Lett.} {\bf 83} (1999) 3605--3608,
  [\href{http://arxiv.org/abs/hep-th/9903190}{{\tt hep-th/9903190}}].

\bibitem{Verlinde:1999fy}
H.~L. Verlinde, {\it {Holography and compactification}},  {\em Nucl.Phys.} {\bf
  B580} (2000) 264--274, [\href{http://arxiv.org/abs/hep-th/9906182}{{\tt
  hep-th/9906182}}].

\bibitem{de-Boer:1999xf}
J.~de~Boer, E.~P. Verlinde, and H.~L. Verlinde, {\it {On the holographic
  renormalization group}},  {\em JHEP} {\bf 0008} (2000) 003,
  [\href{http://arxiv.org/abs/hep-th/9912012}{{\tt hep-th/9912012}}].

\bibitem{Verlinde:1999xm}
E.~P. Verlinde and H.~L. Verlinde, {\it {RG flow, gravity and the cosmological
  constant}},  {\em JHEP} {\bf 0005} (2000) 034,
  [\href{http://arxiv.org/abs/hep-th/9912018}{{\tt hep-th/9912018}}].

\bibitem{Balasubramanian:2000wv}
V.~Balasubramanian, E.~G. Gimon, and D.~Minic, {\it {Consistency conditions for
  holographic duality}},  {\em JHEP} {\bf 0005} (2000) 014,
  [\href{http://arxiv.org/abs/hep-th/0003147}{{\tt hep-th/0003147}}].

\bibitem{de-Haro:2000xn}
S.~de~Haro, S.~N. Solodukhin, and K.~Skenderis, {\it {Holographic
  reconstruction of space-time and renormalization in the AdS / CFT
  correspondence}},  {\em Commun.Math.Phys.} {\bf 217} (2001) 595--622,
  [\href{http://arxiv.org/abs/hep-th/0002230}{{\tt hep-th/0002230}}].

\bibitem{Manohar:1983md}
A.~Manohar and H.~Georgi, {\it {Chiral Quarks and the Nonrelativistic Quark
  Model}},  {\em Nucl.Phys.} {\bf B234} (1984) 189.

\bibitem{Georgi:1986kr}
H.~Georgi and L.~Randall, {\it {Flavor Conserving CP Violation in Invisible
  Axion Models}},  {\em Nucl.Phys.} {\bf B276} (1986) 241.

\bibitem{Chacko:1999hg}
Z.~Chacko, M.~A. Luty, and E.~Ponton, {\it {Massive higher dimensional gauge
  fields as messengers of supersymmetry breaking}},  {\em JHEP} {\bf 0007}
  (2000) 036, [\href{http://arxiv.org/abs/hep-ph/9909248}{{\tt
  hep-ph/9909248}}].

\end{thebibliography}\endgroup

\end{document}